\documentclass[a4paper,11pt]{article}
\usepackage{pos}

\usepackage{subfigure}
\usepackage{ragged2e}

\newcommand{\ie}{{\textit{ i.e.},}\ }

\usepackage{amsmath}
\usepackage{empheq}
\usepackage[most]{tcolorbox}

\newtcbox{\mymath}[1][]{%
    nobeforeafter, math upper, tcbox raise base,
    enhanced, colframe=blue!30!black,
    colback=blue!30, boxrule=1pt,
    #1}
    
    \usepackage{dsfont}

\usepackage[framemethod=TikZ]{mdframed}
\usepackage{amsthm}

\newcounter{theo}[section] 
\renewcommand{\thetheo}{\arabic{section}.\arabic{theo}}
\newenvironment{theo}[2][]{%
\refstepcounter{theo}%
\ifstrempty{#1}%
{\mdfsetup{%
frametitle={%
\tikz[baseline=(current bounding box.east),outer sep=0pt]
\node[anchor=east,rectangle,fill=blue!20]
{\strut Theorem~\thetheo};}}
}%
{\mdfsetup{%
frametitle={%
\tikz[baseline=(current bounding box.east),outer sep=0pt]
\node[anchor=east,rectangle,fill=blue!20]
{\strut Box~\thetheo:~#1};}}%
}%
\mdfsetup{innertopmargin=10pt,linecolor=blue!20,%
linewidth=2pt,topline=true,%
frametitleaboveskip=\dimexpr-\ht\strutbox\relax
}
\begin{mdframed}[]\relax%
\label{#2}}{\end{mdframed}}

\title{Modave Lecture Notes on de Sitter Space \& Holography}

\author*[a]{Dami\'an A. Galante}

\affiliation[a]{King's College London, the Strand, London WC2R 2LS, UK}

\emailAdd{damian.galante@kcl.ac.uk}

\abstract{
 These lecture notes provide an overview of different aspects of de Sitter space and their plausible holographic interpretations. We start with a general description of the classical spacetime. We note the existence of a cosmological horizon and its associated thermodynamic quantities, such as the Gibbons-Hawking entropy. We discuss geodesics and shockwave solutions, that might play a role in a holographic description of de Sitter. Finally, we discuss different approaches to quantum theories of de Sitter space, with an emphasis on recent developments in static patch holography.

\Centering

 \vspace{0.5cm}

     Please send comments, typos or corrections to my email address.

}

\FullConference{%
  Modave Summer School in Mathematical Physics (Modave2022)\\
  5-9 September, 2022\\
  Modave, Belgium}

 \tableofcontents

\begin{document}

\maketitle

\section*{Foreword}
\addcontentsline{toc}{section}{Foreword}

\thispagestyle{plain}

These notes are an extended version of the lectures I gave in the XVIII Modave Summer School in Mathematical Physics in September, 2022. They are oriented primarily to PhD students in theoretical physics, who do not necessarily work on gravity or holography.

I was initially asked to talk about ``holography \textit{in} de Sitter space". However, as you can see from the title, the topic has been slightly changed. Despite many recent developments in understanding quantum features of de Sitter (dS) space, we still lack a full framework. Most of these lectures are devoted to explain the reason for this. In that endeavour, I decided to focus on certain peculiar features of de Sitter space and contrast them with their analogous anti-de Sitter and/or black hole versions.

The lectures are divided into six chapters. The first one is mostly introductory and motivational, summarising the vast experimental evidence we have to date of two cosmological periods of accelerated expansion. The second chapter provides an overview of the geometry of dS space at a classical level. The third one deals with thermodynamic properties of the cosmological horizon. The fourth and fifth study two different probes that have been very useful in the context of holography in Anti-de Sitter (AdS): geodesics and shockwave solutions, respectively.

The expert reader can probably skip the first five chapters and move directly to the last one, where I intend to summarise recent developments and proposals for dS holography. The sixth chapter starts by reviewing quantum field theory in a fixed dS background and the dS/CFT correspondence. I then focus on recent ideas regarding static patch holography, including the stretched horizon, a discussion on the role of timelike boundaries, the $T\bar{T} + \Lambda_2$ construction and dS holography in two dimensions. I tried to compile a comprehensive list of recent references on these subjects. But this, of course, can only be a partial selection of topics and references related to quantum aspects of dS space. Other very interesting ones, such as, for instance, inflation, the wavefunction of the Universe and infrared divergences are not discussed here.

During the actual lectures in Modave, I spent considerable time discussing scalar field theory in a fixed dS background. I shortened that discussion in the present notes, considering this has already been properly reviewed in other places. See, for instance, \cite{Spradlin:2001pw, Anninos:2012qw} for excellent reviews on the subject. For classical aspects of the geometry \cite{hartman} provides a nice overview, while \cite{Bousso:2002fq} provides a summary of quantum problems involving the cosmological horizon. There are certainly many other useful reviews on the subject. 

While, of course, these notes may overlap with some of the other reviews at different points, my intention is to provide an updated look into the subject. Special emphasis is given to tools and features that have recently been particularly successful in the context of AdS holography. Hopefully, these will also play some role in a modern understanding of the quantum nature of de Sitter space. 

Hope you enjoy!

\clearpage

\section{Introduction to de Sitter space}
As mathematical physicists, we probably do not need much of a motivation to study either de Sitter or Anti-de Sitter spaces. Together with flat space, these are the three maximally symmetric spacetimes with positive, negative or zero cosmological constant $\Lambda$, and as such, they provide a rich mathematical structure to discuss both classical and quantum foundational issues in gravity. In fact, textbooks like \cite{hawking_ellis_1973} already provide a basic treatment of both (A)dS.

However, the role of both (A)dS changed dramatically, and for different reasons, in 1998. On the theoretical side, the first concrete realisation of a holographic picture in gravity was derived for asymptotically AdS spacetimes \cite{Maldacena:1997re}. On the experimental side, astrophysical observations of supernovae \cite{SupernovaSearchTeam:1998fmf, SupernovaCosmologyProject:1998vns} showed that the Universe is currently expanding at an accelerated rate, indicating that our cosmological constant is small, but positive.

\

\begin{figure}[h!]
        \centering
        \subfigure[Cosmic Microwave Background]{
                \includegraphics[height=4.5cm]{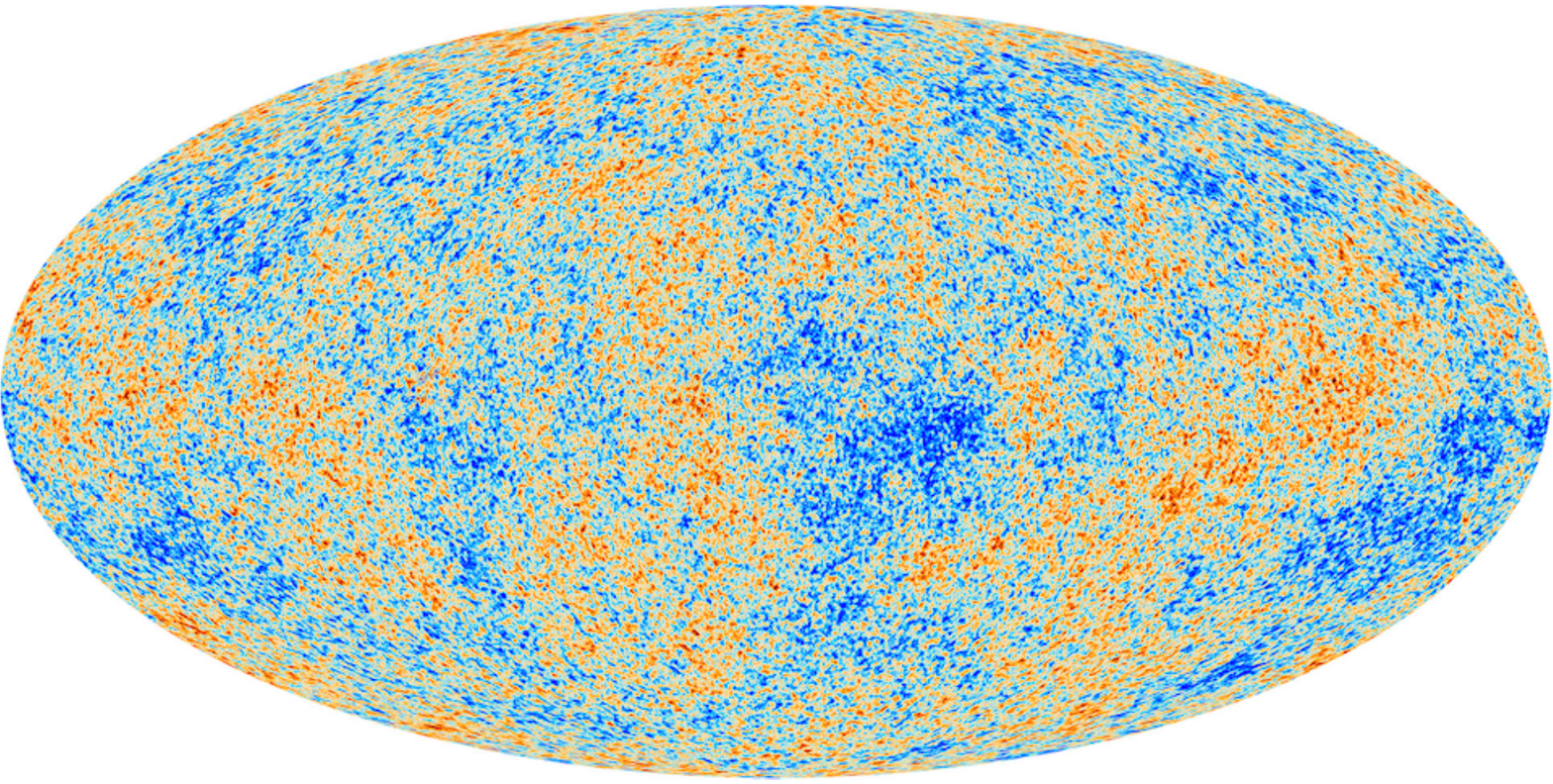} \label{fig:planck}} \qquad \qquad
         \subfigure[Supernovae]{
                \includegraphics[height=4.5cm]{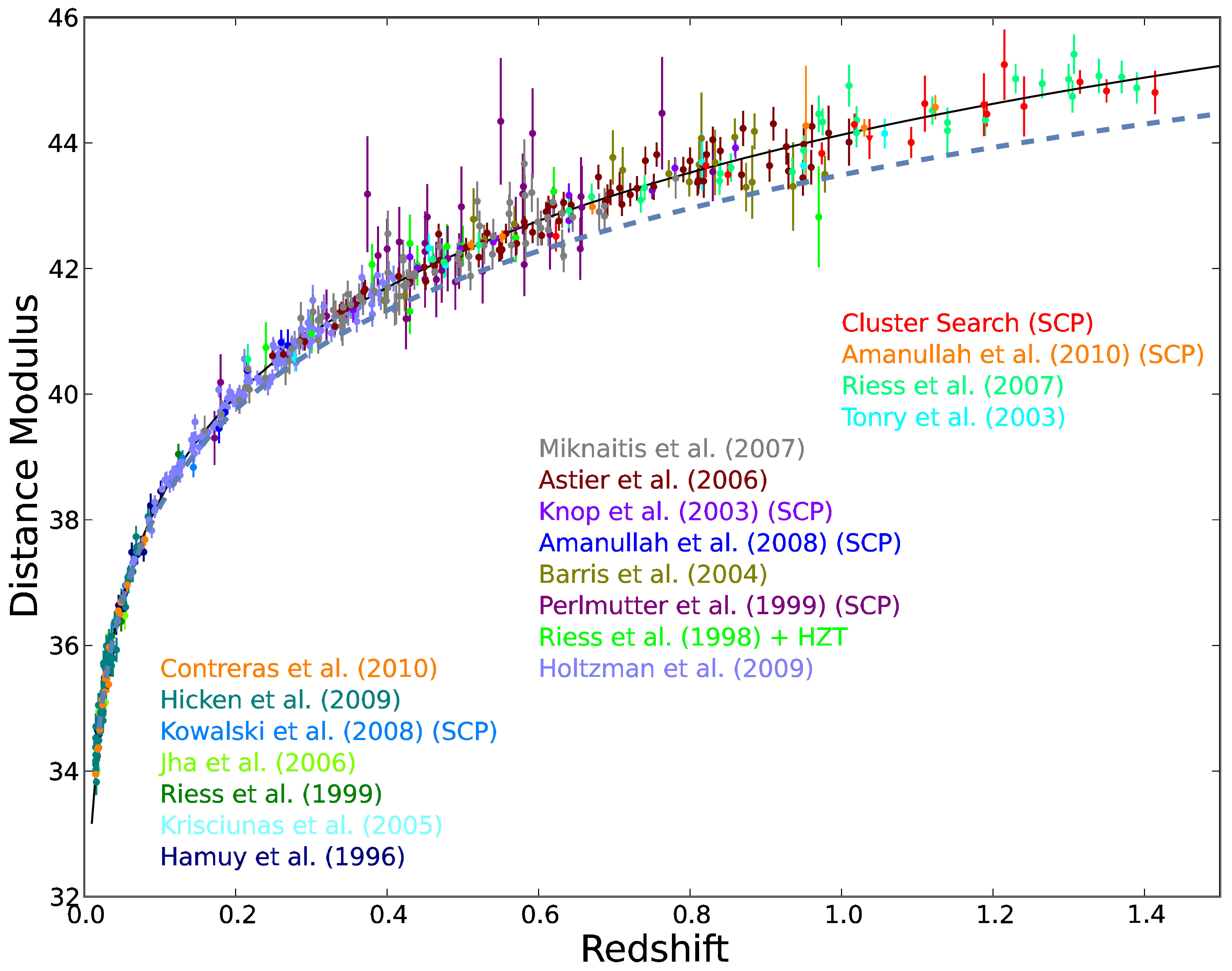}\label{fig:super}} 
                 \caption{\footnotesize (a) Temperature fluctuations of the CMB, taken from \cite{refId0}. Measurements are consistent with cosmic inflationary models. (b) Hubble diagram for the Union 2.1 compilation of type Ia supernovae. The solid line is the best-fit of the data, consistent with a $\Lambda$CDM flat cosmology with $\Omega_\Lambda \sim 0.7$. Figure taken from \cite{Suzuki_2012}. In dashed blue, we added for comparison the prediction obtained for a flat Universe with no cosmological constant.}
\end{figure}

\noindent \textbf{Observational evidence.} We also have evidence that our Universe underwent a period of accelerated expansion at the very beginning of time. Experiments like COBE detected a mostly isotropic \textbf{Cosmic Microwave Background} (CMB) in the form of black body radiation at a temperature of $T=2.73K$, with relative fluctuations of the order of $10^{-5}$. But it was only in 2003 that the WMAP experiment managed to measure a nearly scale-invariant spectrum, that was consistent with cosmic inflation models. This provided supporting evidence for this theory of accelerated expansion during the first instances of the Universe. See, for instance, \cite{baumann_2022}. The Planck satellite even improved this measurement in 2013, see figure \ref{fig:planck}. 

\begin{figure}[h!]
        \centering
        \includegraphics[scale=0.5]{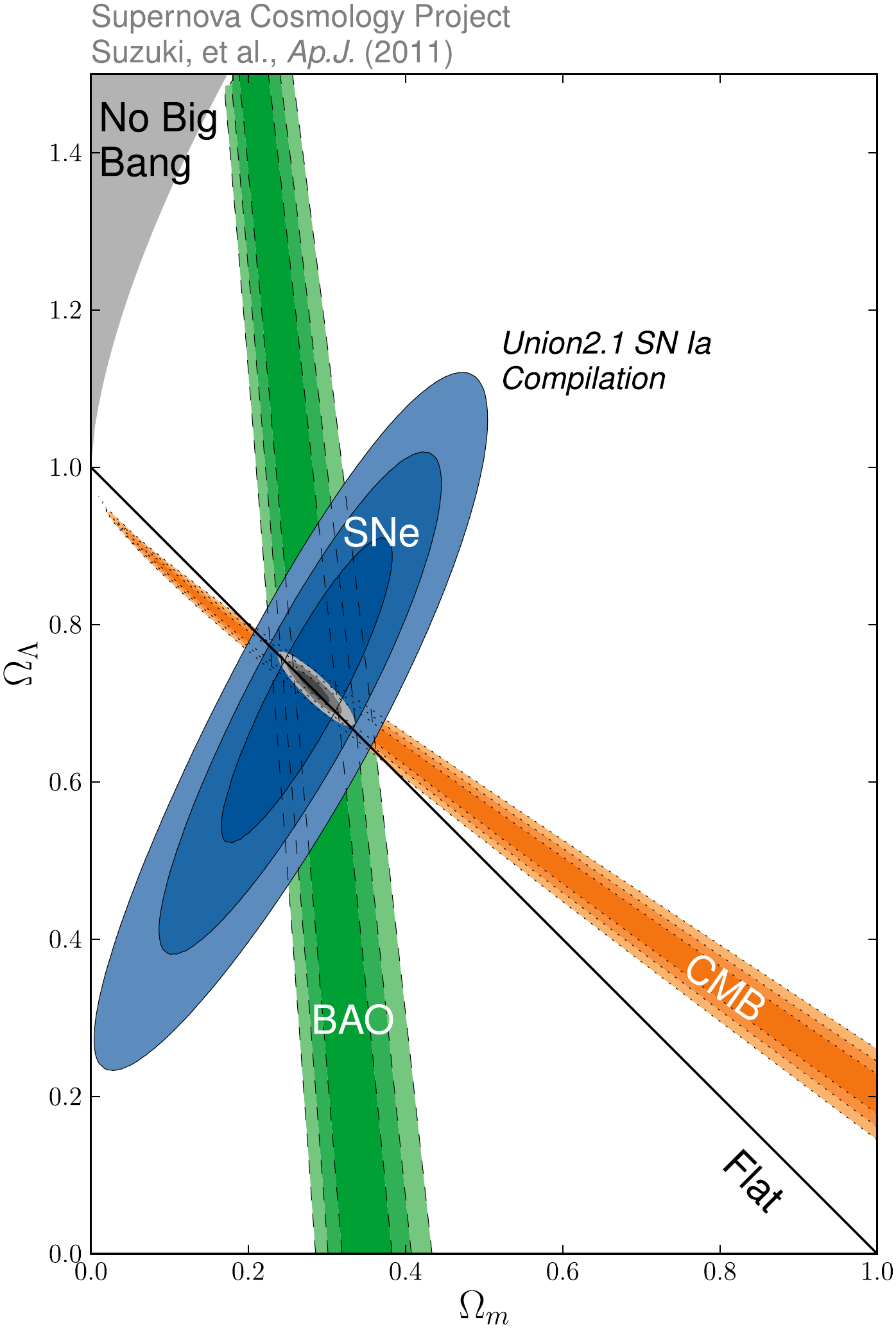}
\caption{$\Lambda$CDM model: $68.3\%, 95.4\%,$ and $99.7\%$ confidence regions of the $(\Omega_m,\Omega_\Lambda)$ plane from SNe Ia combined with the constraints from BAO and CMB. Figure from \cite{Suzuki_2012}.} \label{fig:three}
\end{figure}

Apart from the inflationary era, we now have at least three different types of experiments supporting a current cosmological era of accelerated expansion.  The first one is the already mentioned measurement of \textbf{supernovae} (SNe). The data from 1998 has been updated with more than 500 observations from 19 different datasets that show that large redshift supernovae appear farther away than they should if there were no accelerated expansion \cite{Suzuki_2012}. See figure \ref{fig:super}. The data from the CMB can also be used to constrain the value of the cosmological constant. And finally, there are measurements from \textbf{Baryon Acoustic Oscillations} (BAO). BAOs are fluctuations in the density of baryonic matter that can be seen in our Universe and were caused by acoustic waves in the primordial plasma of the early Universe. They were first observed by the Sloan Digital Sky Survey \cite{sdss} and the 2dF Galaxy Redshift Survey \cite{osci} in 2005 and by comparing with CMB data, they provide another measurement of the cosmological constant. Neither of these three alone provides definite certainty of a positive cosmological constant, but combined together they constitute fairly convincing evidence that we currently live in a spatially flat Universe dominated by a positive cosmological constant with $\Omega_\Lambda \sim 0.7$ \cite{Suzuki_2012}, see figure \ref{fig:three}. Challenges, anomalies and extensions to this cosmological model (including the Hubble tension) are reviewed in \cite{Perivolaropoulos:2021jda, Abdalla:2022yfr, DES:2022ccp}.

\

\noindent \textbf{A connection to holography.} Even if our Universe is not exactly dS at the moment, cosmic no hair theorems \cite{Wald:1983ky} predict that in (most) cosmological scenarios, as time evolves, all other matter and energy content in the Universe will dilute and we will be asymptotically approaching a locally dS geometry. Current measures give a very small value of the cosmological constant,
\begin{equation}
    \Lambda \sim 10^{-52} m^{-2} \sim 10^{-122} \ell_P^{-2} \,,
\end{equation}
where $\ell_P$ is the Planck length. This famously differs from the theoretical effective field theory expectation by around 122 orders of magnitude \cite{carroll_2019}, and constitutes what is known as the cosmological constant problem. Understanding quantum features of spacetime might provide a way of addressing this discrepancy. One complication is that it has been hard to realise universes with positive cosmological constant in string theory \cite{Danielsson:2018ztv}.

However, as we will review in section \ref{sec:ds_entropy}, observers in an ever-accelerating spacetime are surrounded by cosmological event horizons. It has been proposed that the cosmological horizon in four-dimensional dS carries an entropy given by \cite{Gibbons:1977mu, Gibbons:1976ue}
\begin{empheq}[box={\mymath[drop lifted shadow]}]{equation}
    S = \frac{3\pi}{\Lambda} \frac{k_B c^3}{\hbar G_N} \,, \label{dS_entropy}
\end{empheq}
that if interpreted in the statistical mechanics sense, would bind the value of the cosmological constant to the microscopic structure of the Universe. Moreover, the fact that this formula is an area entropy formula (as in the black hole case) also points towards a holographic principle for this type of spacetime. In fact, the traditional arguments towards the realisation of holography in gravity do not rely on the value or sign of the cosmological constant \cite{tHooft:1993dmi, Susskind:1994vu}. A holographic description of dS space is then highly desirable. But let us start from the very beginning.

\section{The basics of de Sitter space} \label{ch:what_is}
\begin{figure}[h!]
        \centering
        \includegraphics[height=6cm]{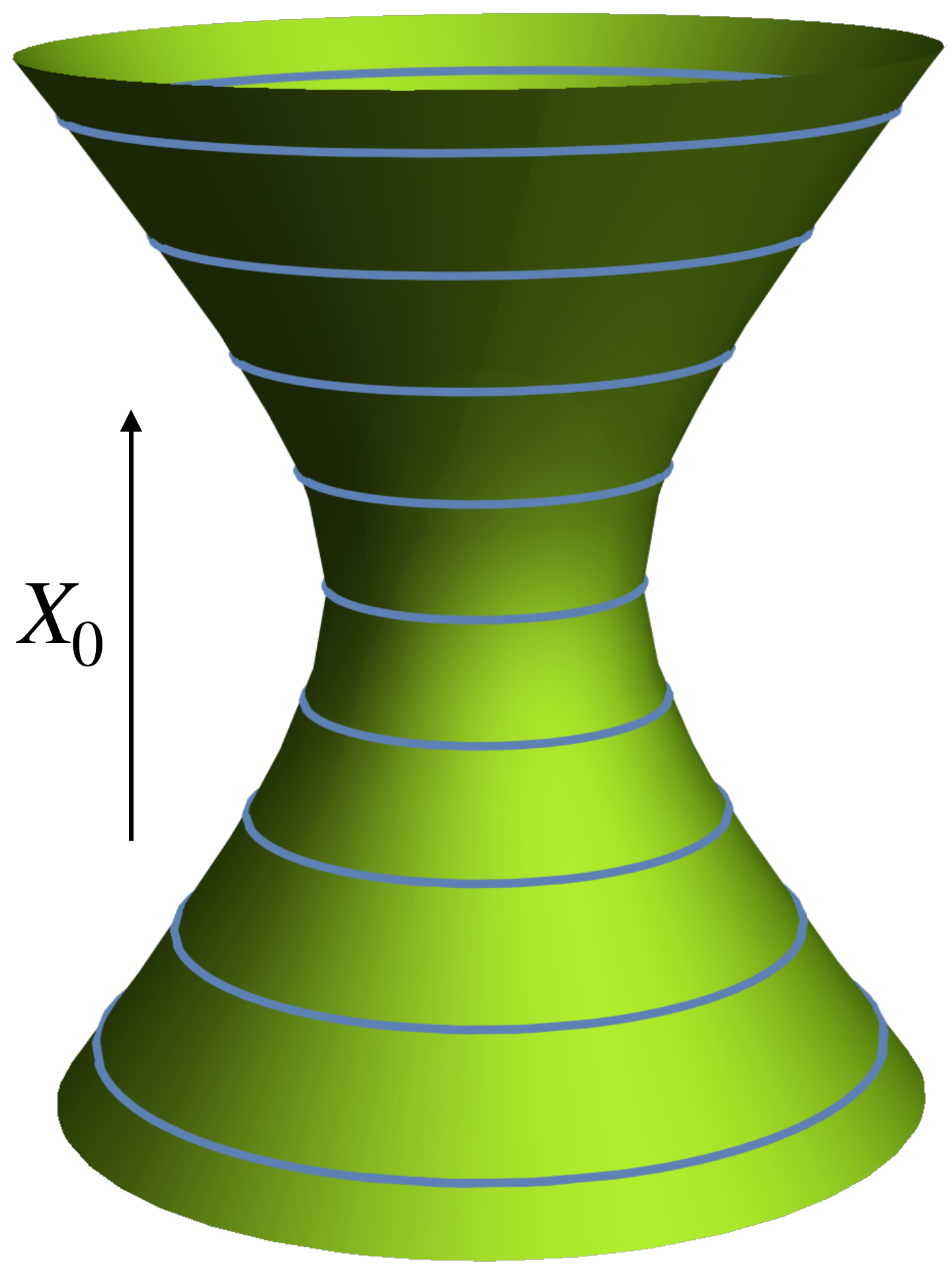}
\caption{De Sitter in 2d as the hyperboloid embedded in 3d Minkowski space. Constant $X_0$ slices are circles with varying radius, some of which are drawn in blue.}\label{fig:hyperboloid}
\end{figure}

We will first review the classical geometry of dS spacetime. Most of this section can also be read in, for instance, \cite{Spradlin:2001pw, Anninos:2012qw}. The easiest way to visualise de Sitter (dS) space is through the embedding picture. Consider Minkowski spacetime in $(d+1)$ dimensions, $\mathcal{M}^{d+1}$. In our conventions, the Minkowski metric is given by
\begin{equation}
    ds^2_{\mathcal{M}^{d+1}} = -dX_0^2 + dX_1^2 + \cdots + dX_d^2 \,. \label{mink metric}
\end{equation}
Then dS space in $d$ dimensions is realised as the following hypersurface embedded in $\mathcal{M}^{d+1}$,
\begin{empheq}[box={\mymath[drop lifted shadow]}]{equation}
-X_0^2 + X_1^2 + \cdots + X_d^2 = \ell^2 \,, \label{hyperboloid}
\end{empheq}
where $\ell$ is called the curvature scale or the dS radius. It is easy to see that this equation defines a hyperboloid in $\mathcal{M}^{d+1}$, shown in figure \ref{fig:hyperboloid}. It is also straightforward to realise that the set of coordinate transformations that leave (\ref{hyperboloid}) unchanged is given by the $SO(d,1)$ group, which is then the group of isometries of dS space in $d$ dimensions. Note that this is the Euclidean conformal group in $(d-1)$ dimensions.

De Sitter space is the maximally symmetric Einstein manifold with positive curvature, so it satisfies 
\begin{equation}
    R_{\mu\nu} - \frac{1}{2} R g_{\mu\nu} + \Lambda g_{\mu\nu} = 0 \,, \qquad \text{with} \qquad  \Lambda = \frac{(d-2)(d-1)}{2\ell^2} \,,
    \label{einstein_eqs}
\end{equation}
where $\Lambda$ is a \textit{positive} cosmological constant and $d>2$. In $d=2$ dilaton-gravity theories, the cosmological constant is usually set to $\Lambda = \ell^{-2}$. For most of these lectures we will work in general spacetime dimension $d$. Though the case of $d=4$ is, of course, the most relevant to our Universe, we will sometimes go to $d=2,3$ for simplicity. 

\begin{theo}[Anti de Sitter]{theo:theo2}
Anti-de Sitter (AdS) can also be viewed as a hyperboloid embedded in a higher dimensional manifold, but a different one. AdS in $d$ dimensions is given by
\begin{equation}
-X_0^2 -X_d^2 + X_1^2 + \cdots + X_{d-1}^2 = - R^2 \,,
\end{equation}
where now $R$ is the AdS radius and this surface is embedded in $\mathbb{R}^{2,d-1}$, \ie $ds^2 = -dX_0^2 -dX_d^2 +dX_1^2 + \cdots + dX_{d-1}^2$. Famously, the AdS isometries form the group $SO(d-1,2)$, that is the conformal group in $(d-1)$ dimensions and the first hint towards the AdS/CFT correspondence.
\end{theo}

In what follows, we will describe dS with different coordinate systems that cover different parts of the whole hyperboloid.
\subsection{Coordinate systems}
For now, we will set $\ell = 1$. It will be convenient to define coordinates $\omega^i$ on the unit sphere $S^{d-1}$, such that $\sum_{i=1}^d (\omega^i)^2 = 1$. When in $d=2,3$ we consider a spatial circle, we choose an angular coordinate $\varphi \in (0,2\pi]$ to parameterise it.

\subsubsection{Global coordinates}
The first coordinates that we will study cover the full hyperboloid and are called global coordinates, $\{\tau, \omega^i\}$. The coordinate $\tau$ is usually called the global time and the embedding is given by
\begin{eqnarray}
\begin{cases}
X^0 & =  \sinh \tau \,, \\
X^i & =  \cosh \tau \, \omega^i \,.
\end{cases}
\end{eqnarray}
It is straightforward to verify that this satisfies (\ref{hyperboloid}) with $\ell=1$. Plugging this into (\ref{mink metric}), we obtain the induced metric on the hyperboloid, or the global dS metric,
\begin{empheq}[box={\mymath[drop lifted shadow]}]{equation}
    ds^2 = -d\tau^2 + \cosh^2 \tau \, d\Omega_{d-1}^2 \,,
    \label{global metric}
\end{empheq}
where $d\Omega_{d-1}^2$ is the metric on the $S^{d-1}$ and $\tau \in [-\infty, \infty]$. Constant time slices are then compact. For $\tau>0$, this is the typical picture of a closed Universe whose size is expanding exponentially as time evolves forward. The minimal size of the sphere is at $\tau = 0$, where the radius of the sphere is one (in units of the dS radius). Note that this metric depends explicitly on the global time; dS does not have a global timelike Killing vector. We will explore some consequences of this later on.

\subsubsection{Conformal coordinates and Penrose diagram}
\begin{figure}[h!]
        \centering
        \includegraphics[height=6cm]{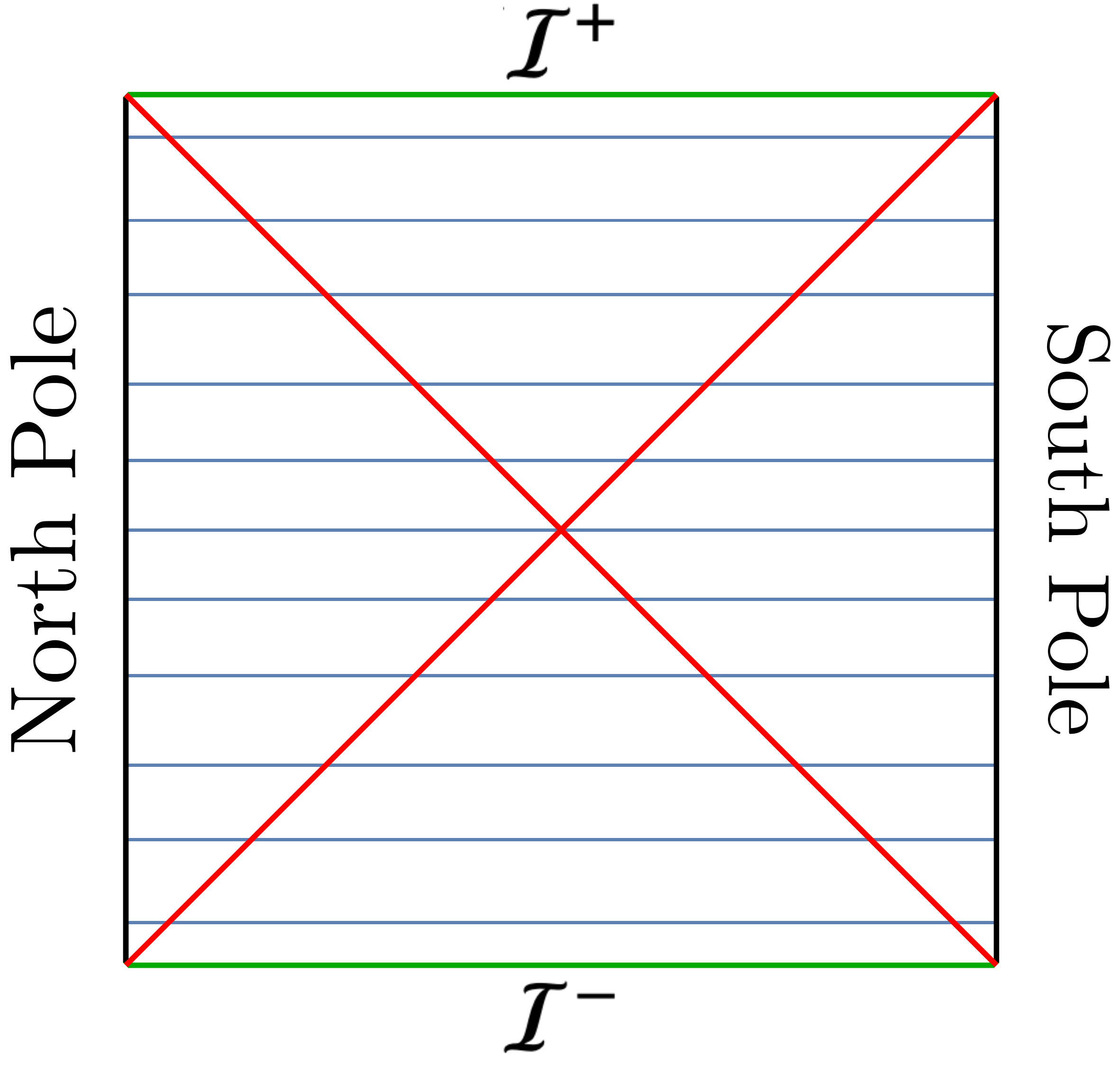}
\caption{Penrose diagram of de Sitter space.}\label{fig:penrose}
\end{figure}

To look at the causal structure of dS space, it is convenient to define a conformal time $T$ such that $\cos T = \cosh^{-1} \tau$. From here, it follows that $-\pi/2 \leq T \leq \pi/2$ and the metric becomes,
\begin{equation}
ds^2 = \frac{1}{\cos^2 T} \left( -dT^2 + d\theta^2 + \cos^2 \theta d\Omega_{d-2}^2 \right) \,.
\end{equation}
This metric is useful to find the Penrose diagram as null rays in the $\theta$-direction travel at 45 degrees angles. The coordinate $\theta$ now spans from $-\pi/2$ to $\pi/2$. In the Penrose diagram, we only draw the $T$ and $\theta$ coordinates, so it is quite clear that the Penrose diagram of dS space becomes a square, as appears in figure \ref{fig:penrose}. 

Each horizontal line in the diagram (in blue) corresponds to a $(d-1)$ sphere, whose radius is given by $\cos^{-1}T$. Each point in each line corresponds to a $(d-2)$ sphere with the exception of both vertical edges, which correspond to $\theta = \pm \pi/2$, and thus are not spheres, but single points. We usually call those points the North and South pole of the sphere, and we like to think about inertial observers sitting at those points.

All null rays start (and end) at the infinite past (future) of the dS space, that we call $\mathcal{I}^-$ ($\mathcal{I}^+$). These conformal boundaries are the lower and upper borders of the Penrose diagram (in green) and correspond to slices of infinite size. Observers take infinite proper time to reach $\mathcal{I}^+$.

\begin{theo}[Double sided AdS black hole]{prf:adsbh}
Note that the dS Penrose diagram is identical to the Penrose diagram of the double sided AdS$_d$ black hole (at least for $d=2,3$), whose metric in $d=3$ is given by
\begin{equation}
    ds^2 = -(r^2 - r_h^2) dt^2 + \frac{dr^2}{(r^2-r_h^2)} + r^2 d\varphi^2 \,.
    \label{bh_metric}
\end{equation}
The black hole horizon is at $r = r_h$. However, the two spacetimes are quite different. In particular, if we look at the size of the circle in the AdS case, it grows towards the boundary at $r\to \infty$, while in the dS case, it shrinks to zero size. Sometimes people like to include small  arrows in the direction of growth of the compact sphere to distinguish the two diagrams. The double-sided AdS black hole plays a prominent role in the AdS/CFT correspondence as it is identified as dual to the thermofield double state in the boundary theory \cite{Israel:1976ur,Maldacena:2001kr}.
\end{theo}

\subsubsection{Static coordinates}
A very important aspect of dS space is that no single observer has access to the full spacetime. This is clear by just looking at the Penrose diagram. For instance, a lightray emerging from the North pole at $\mathcal{I}^-$, will only reach the South pole in the infinite future $\mathcal{I}^+$.

\begin{figure}[h!]
        \centering
        \subfigure[]{
                \includegraphics[scale=0.36]{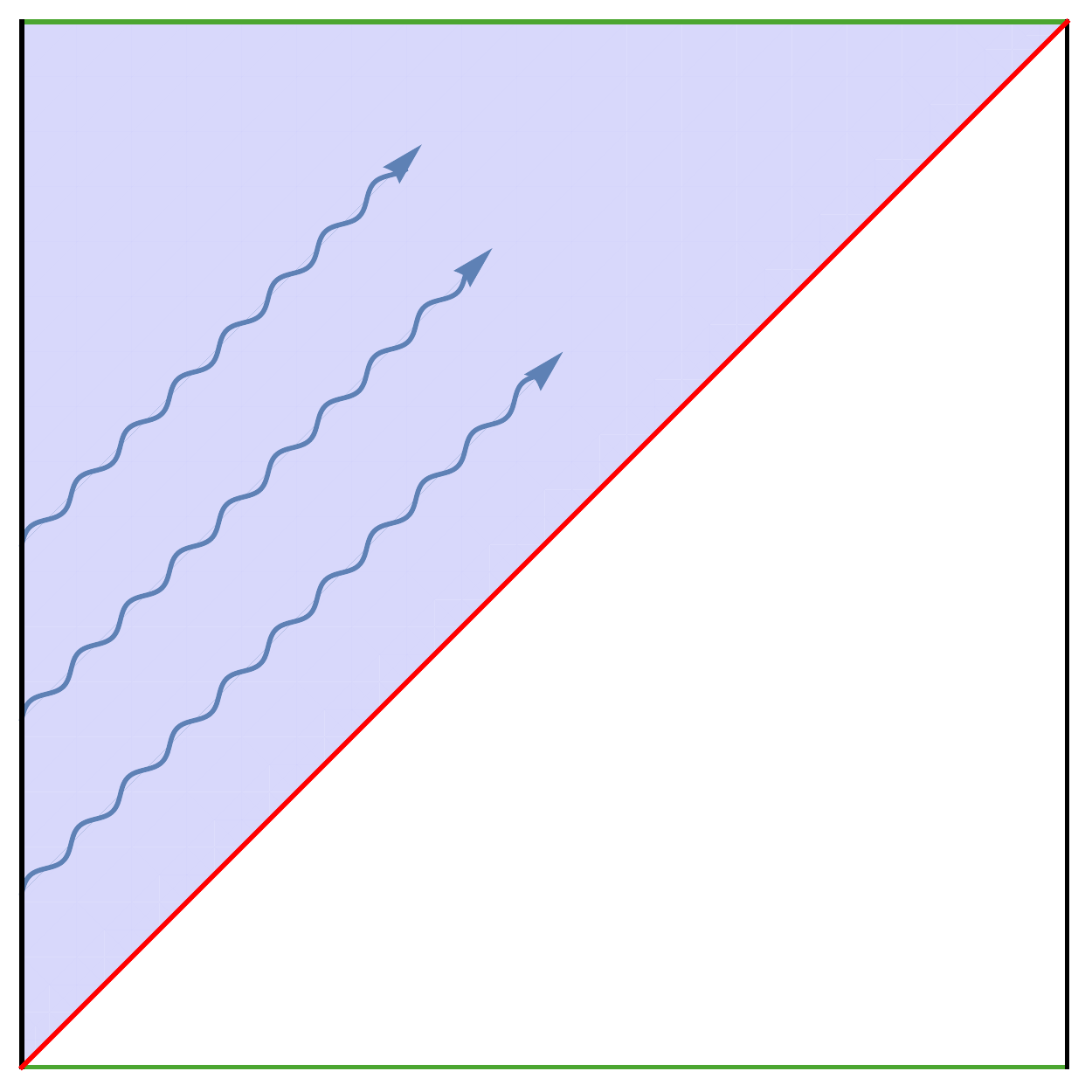} \label{fig:static_1}} \, 
         \subfigure[]{
                \includegraphics[scale=0.36]{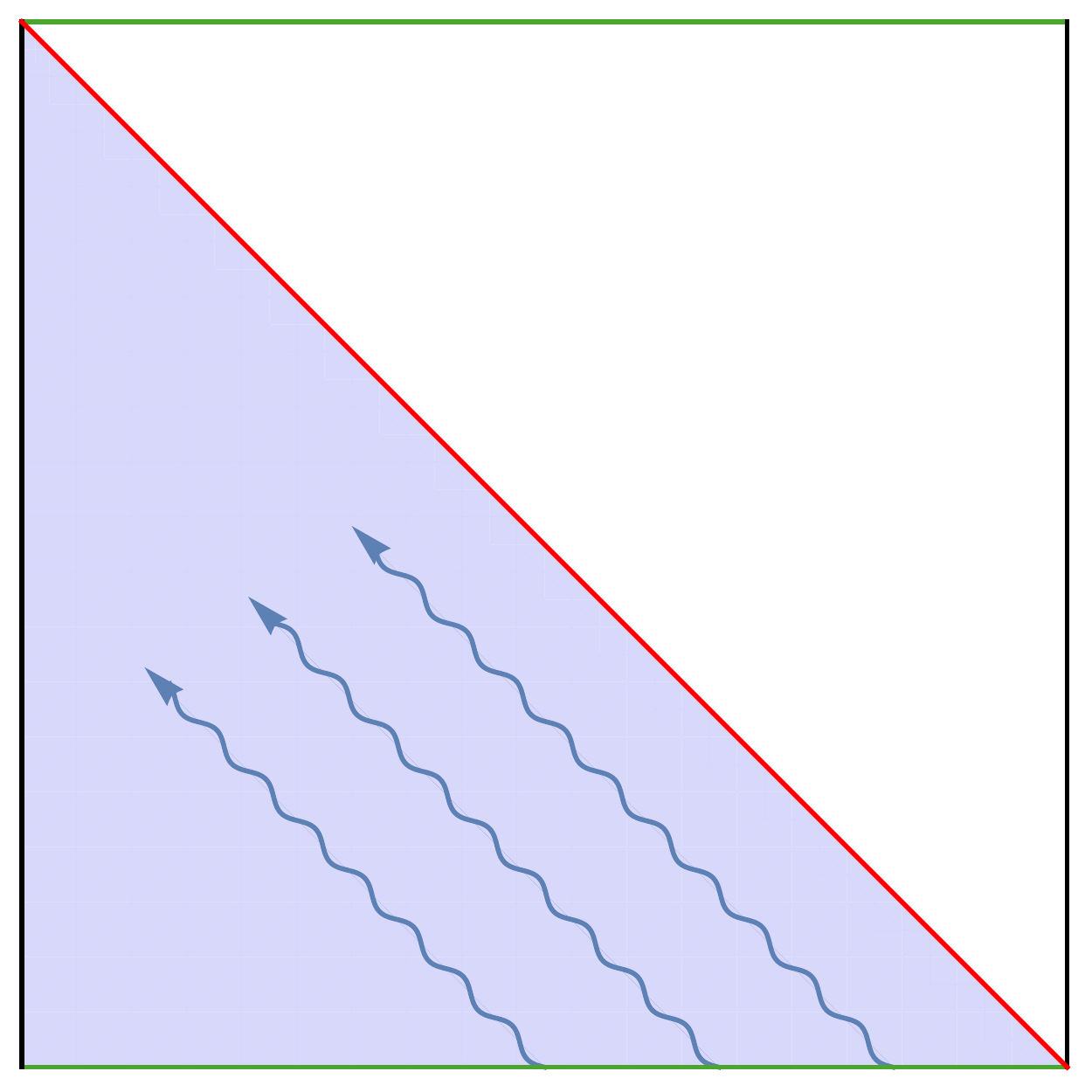}\label{fig:static_2}} \,
         \subfigure[]{
                \includegraphics[scale=0.36]{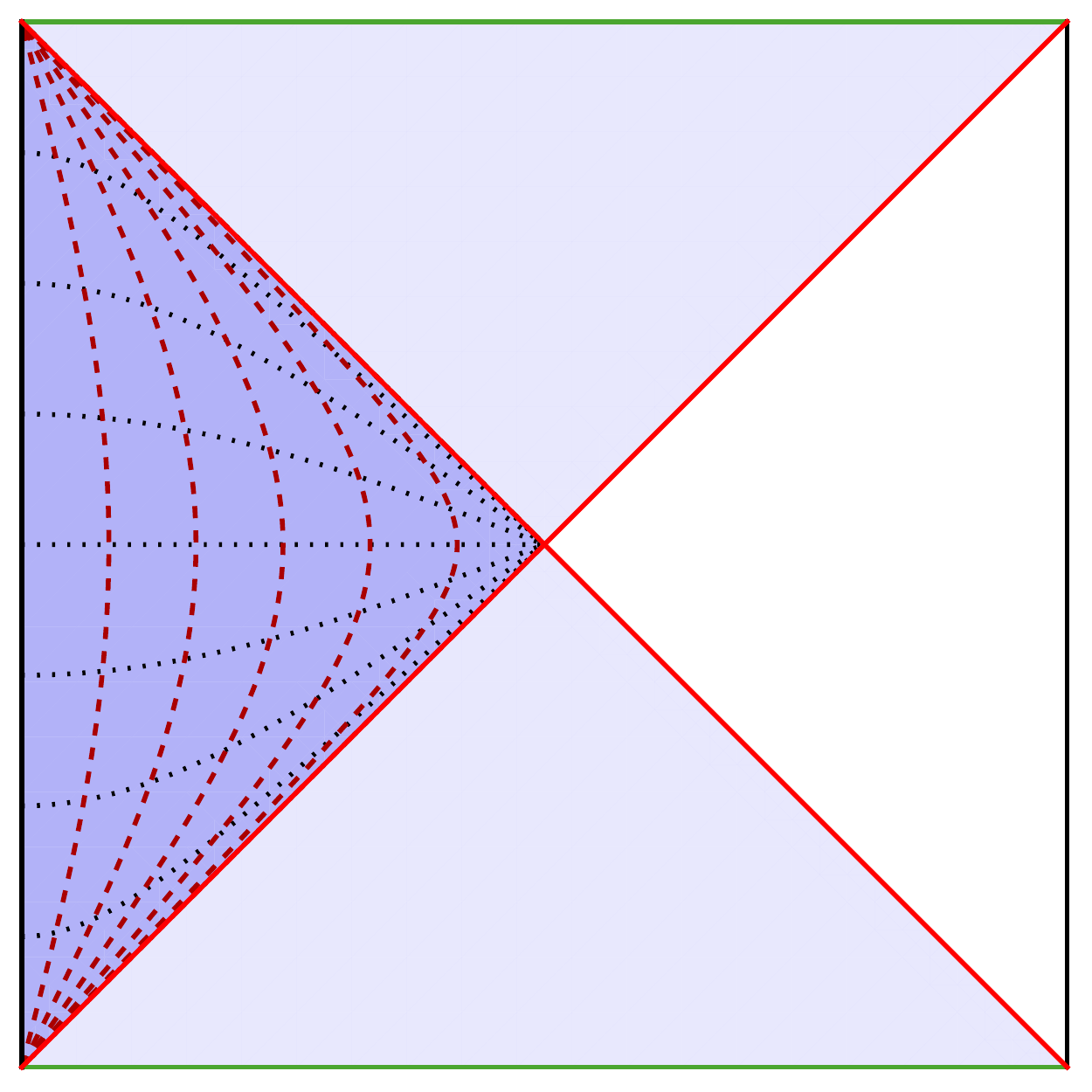}\label{fig:static_3}} 
                 \caption{\footnotesize The intersection between the regions that can be affected by and affect an observer is called the static patch. In \ref{fig:static_3}, we show constant $t$ slices in dotted black and constant $r$ slices in dashed brown.} \label{fig:static}
\end{figure}

An important set of coordinates are those that describe the region accessible to a single observer. This is the intersection between the region of space that can affect the observer and the region that can be affected by them, see figure \ref{fig:static}.

In terms of embedding coordinates, it is described by
\begin{eqnarray}
\begin{cases}
X^0 & =  \sqrt{1-r^2} \sinh t \,, \\
X^i & =  r \omega^i \,, \\
X^d & = \sqrt{1-r^2} \cosh t \,,
\end{cases}
\end{eqnarray}
where $i= 1, \cdots, d-1$. The resulting metric is the so-called static patch metric,
\begin{empheq}[box={\mymath[drop lifted shadow]}]{equation}
    ds^2 = - (1-r^2) dt^2 + \frac{dr^2}{1-r^2} + r^2 d\Omega_{d-2}^2 \,,
    \label{static_metric}
\end{empheq}
where $r \in [0,1]$. There are a number of important observations about this metric:
\begin{itemize}
    \item As the name suggests, this is a static metric; there is no explicit time dependence on the metric. Thus, $\partial_t$ is a timelike Killing vector.
    \item At $r=1$ ($\ell$, if we reinsert the dS length) the norm of the timelike Killing vector vanishes. The $r=1$ surface is a null surface that surrounds the observer at all times. This is what we call the \textbf{cosmological event horizon}. 
    \item We can continue the coordinates for $r > 1$, where the Killing vector becomes spacelike. The situation is similar to an inside out black hole.
\end{itemize}

It is important to note that the existence of this cosmological horizon is a direct consequence of the accelerated expansion of dS space and the finite propagation of the speed of light. There are no singularities or matter in empty dS space but there is still a cosmological horizon. One of the aims of these lectures is to discuss similarities and differences with the usual black hole horizon.

Note that every inertial observer in dS is surrounded by a cosmological horizon. In this sense, it is said that the cosmological horizon is observer dependent, as opposed to the black hole case. It is also the case that observers cannot get rid of their cosmological horizon, making semiclassical processes such as the horizon evaporation extremely subtle \cite{GINSPARG1983245}.

Another difference with the black hole horizon is that the region inside the static patch remains always finite. Comparing (\ref{bh_metric}) to (\ref{static_metric}), we see that the size of the compact space grows to infinite size in the black hole case, while in the static patch geometry it never gets larger than the dS length. As we will discuss later, this imposes severe constraints on how to define observables in the static patch, as usually in gravity we make use of an asymptotic boundary (such as the null boundary of asymptotically flat spacetimes or the timelike boundary of asymptotically AdS spacetimes).

\subsubsection{Other coordinates} \label{others}
There are other sets of coordinates that are useful for different purposes. Here, we just point out some other coordinates that will appear throughout these lecture notes, and refer the reader to \cite{Spradlin:2001pw} in order to obtain them from the embedding picture.

\begin{figure}[h!]
        \centering
        \subfigure[Planar coordinates]{
                \includegraphics[scale=0.36]{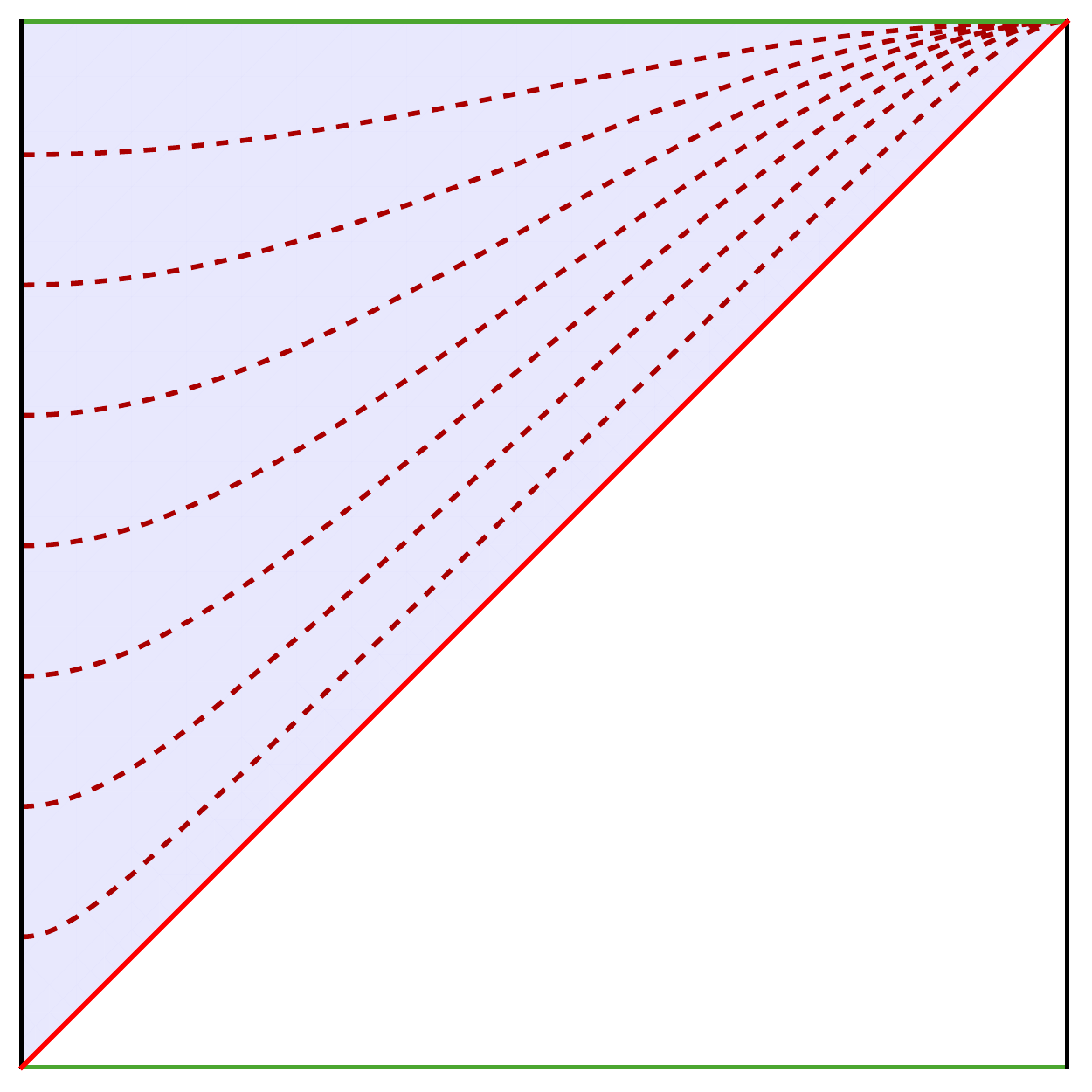} \label{fig:planar}} \qquad \qquad
         \subfigure[dS slicing]{
                \includegraphics[scale=0.36]{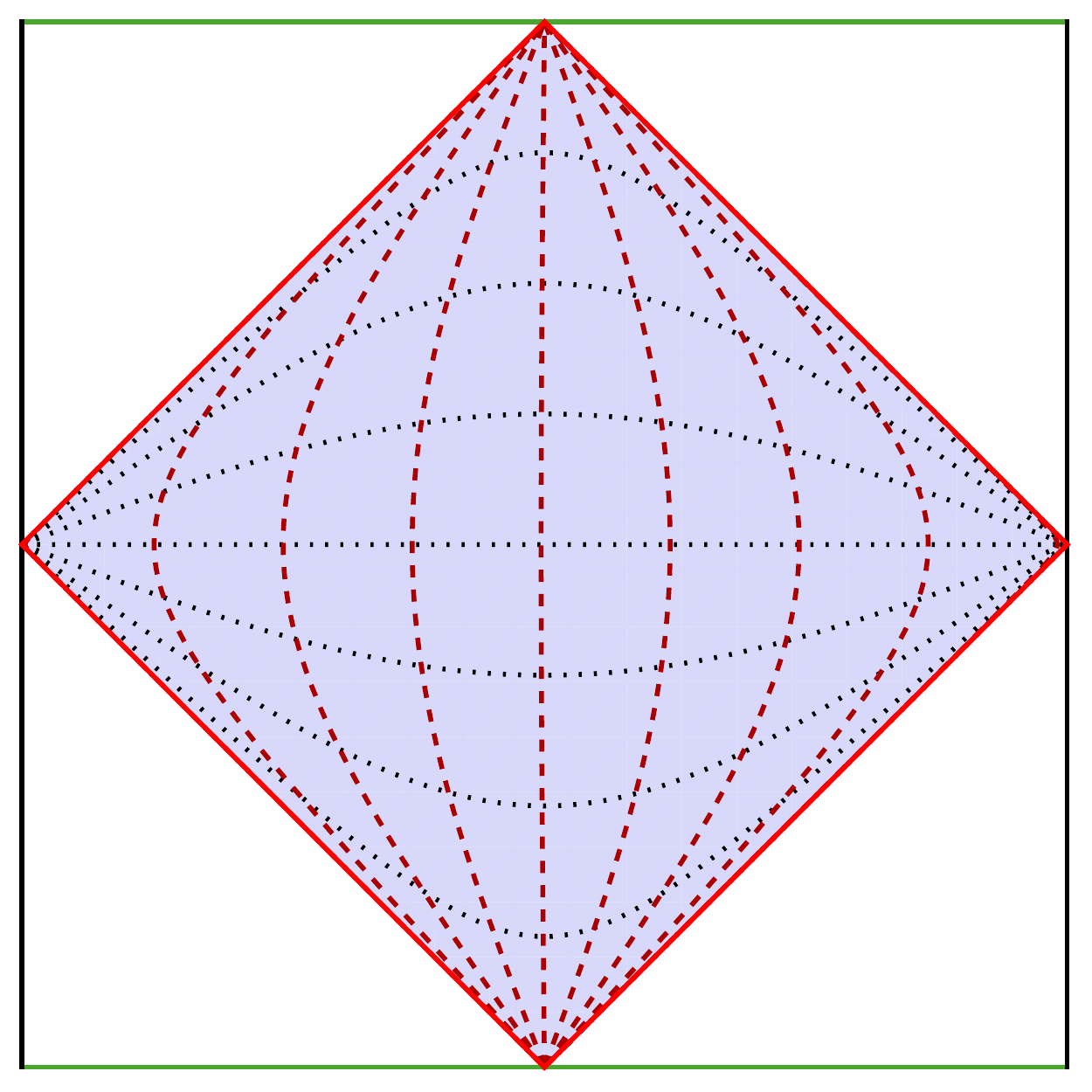}\label{fig:dSdS}} 
                 \caption{\footnotesize De Sitter Penrose diagrams and the patches covered by planar and dS/dS coordinates. In \ref{fig:planar}, we show some constant $\eta$ slices in dashed brown. In \ref{fig:dSdS}, dashed brown curves are constant $\tilde{\omega}$ slices while dotted black lines are constant $\tilde{\tau}$ slices.} \label{fig:penrose2}
\end{figure}

First, we introduce the \textbf{planar coordinate} system, that covers half of the Penrose diagram, as shown in figure \ref{fig:planar}, and is given by
\begin{equation}
    ds^2 = \frac{-d\eta^2 + d\textbf{x}_{d-1}^2}{\eta^2} \,,
\end{equation}
where $\eta$ is usually called the conformal time. This is usually the preferred frame for the computation of cosmological correlators \cite{Baumann:2022jpr}. Note the similarities with the AdS Poincar\'e patch. In fact, one can obtain the Euclidean AdS Poincar\'e patch by analytic continuation of the conformal time and the dS length. This can be useful to relate certain computations in Euclidean AdS to dS. See, for instance, \cite{Sleight:2019mgd, Sleight:2019hfp, Sleight:2020obc}. 

Another set of coordinates, which will be useful to study shockwaves later on, are the \textbf{Kruskal coordinates}. These cover the full Penrose diagram and the metric is given by
\begin{equation}
    ds^2 = \frac{1}{(1- UV)^2} \left(-4 dUdV + (1+UV)^2 d\Omega_{d-2}^2 \right) \,, \label{kruskal_coords}
\end{equation}
where $UV \in [-1,1]$ are null coordinates. The horizons are at $UV = 0$, past and future infinities are at $UV = 1$ and the North and South poles are at $UV = -1$.

\

To finish this section, we present the \textbf{dS/dS patch}, where we foliate dS$_d$ space with dS$_{d-1}$ slices. This is useful for a proposal regarding dS holography that we will briefly discuss in the last chapter. For now, the metric is given by,
\begin{equation}
ds^2 = d\tilde{\omega}^2 + \sin^2 \tilde{\omega} \left( -d\tilde{\tau}^2 + \cosh^2 \tilde{\tau} d\Omega_{d-2}^2 \right) \,,
\end{equation}
where $\tilde{\omega} \in [0,\pi]$ and $\tilde{\tau} \in \mathbb{R}$. This set of coordinates covers the central diamond of the Penrose diagram, as shown in figure \ref{fig:dSdS}, ending at horizons for $\tilde{\omega} = 0,\pi$.

\subsection{Euclidean de Sitter space - the sphere} \label{sec:sphere}

In the next chapter, we will study some properties of Euclidean de Sitter space. Consider dS space in global coordinates. Analytically continuing $\tau \to - i \tau_E$ takes the metric in (\ref{global metric}) to
\begin{equation}
    ds^2 = d\tau_E^2 + \cos^2 \tau_E \, d\Omega_{d-1}^2 \,, \label{global_sphere}
\end{equation}
which is the round metric on the $d$-dimensional sphere, $S^d$. It is interesting to note that the analytic continuation of the static patch time, $t \to - i t_E$ in (\ref{static_metric}), also takes you to $S^d$, in a different foliation,
\begin{equation}
      ds^2 = (1-r^2) dt_E^2 + \frac{dr^2}{1-r^2} + r^2 d\Omega_{d-2}^2 \,.
\end{equation}
The same is true for the dS/dS patch upon taking $\tilde{\tau} \to -i \tilde{\tau_E}$, so the sphere plays a predominant role when using Euclidean techniques to study dS space.

\subsection{Schwarzschild de Sitter black holes} \label{sec:sds}

There are also black hole solutions to the Einstein equations with positive cosmological constant. The simplest of them is the Schwarzschild de Sitter (SdS) black hole, that in 4 dimensions is given by the metric element
\begin{equation}
    ds^2 = - \left(1-r^2 - \frac{2M}{r} \right) dt^2 + \frac{dr^2}{\left(1-r^2 - \frac{2M}{r} \right)} + r^2 d\Omega_{2}^2 \,.
    \label{sds metric}
\end{equation}

\begin{figure}[h!]
        \centering
        \includegraphics[height=4cm]{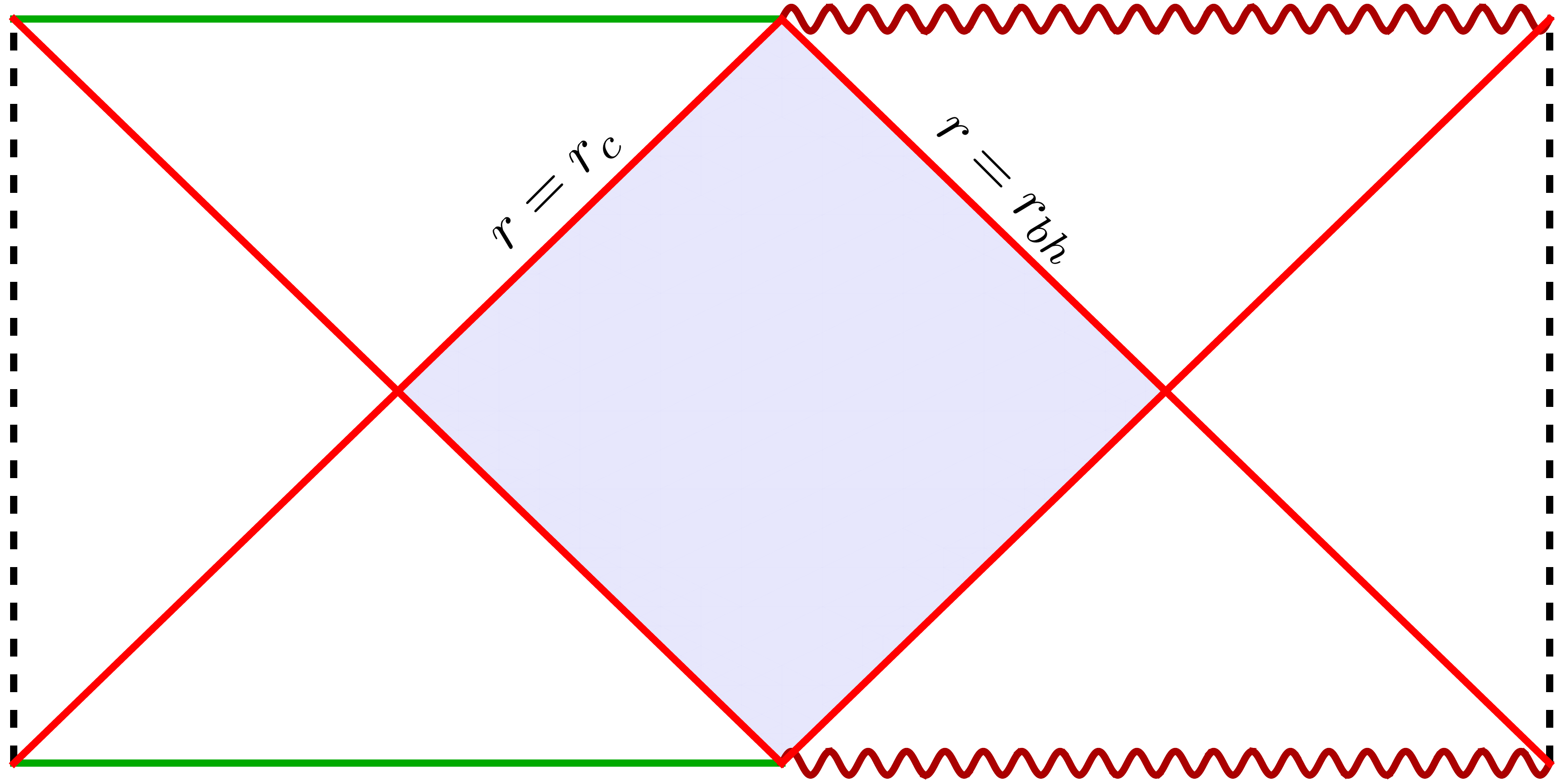}
\caption{Penrose diagram of a Schwarzschild de Sitter black hole. The region between the cosmological $r_c$ and the black hole horizon $r_{bh}$ is one static patch of the SdS spacetime. The dashed black lines at the borders can be either identified or continued endlessly.} \label{fig:sch_dS}
\end{figure}

The Penrose diagram for the SdS space can be seen in figure \ref{fig:sch_dS}. 

\begin{figure}[h!]
        \centering
        \includegraphics[scale=0.6]{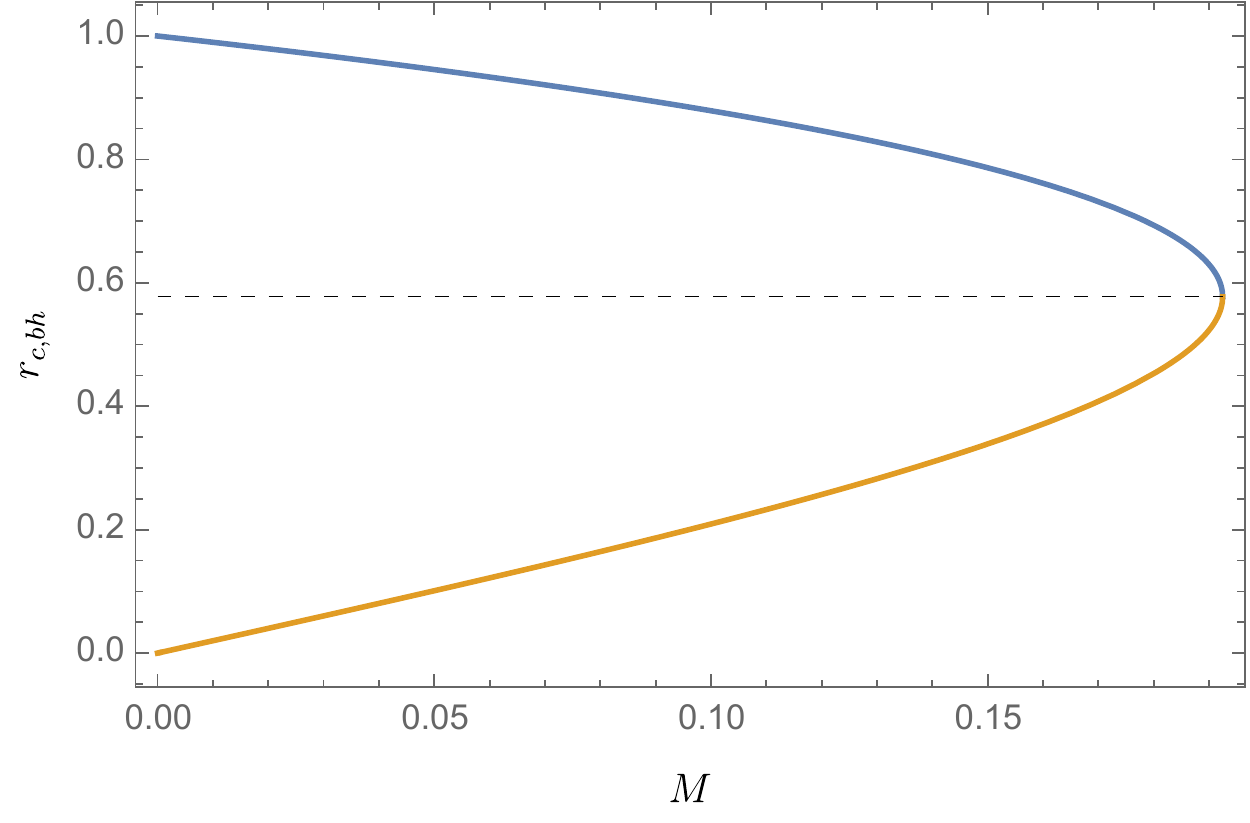}
\caption{Real positive solutions to (\ref{roots}) as functions of $M$. The blue curve corresponds to a cosmological horizon $r_c$, while the yellow one, to a black hole horizon $r_{bh}$. The dashed line in between shows $r_{max}$. We remind the reader that $\ell=1$.}\label{fig:sds_roots}
\end{figure}

The SdS geometry has horizons when the $g_{tt}$ component vanishes, which requires solving the cubic equation
\begin{equation}
    1 - r^2 - \frac{2M}{r} = 0 \,. \label{roots}
\end{equation}
Notice that this equation only has positive real roots for certain values of $M$. For $M=0$, of course, we recover the cosmological horizon of pure dS space. As we increase the value of $M$, we find two positive real solutions that correspond to a cosmological ($r_c$) and a black hole ($r_{bh}$) horizon. The expressions for $r_c$ and $r_{bh}$ can be found analytically and are shown in figure \ref{fig:sds_roots}. Note that $r_c \geq r_{bh}$. As $M$ increases both solutions start getting closer to each other, up until $M= M_{max} \equiv 3^{-3/2}$. The two horizons now have the same radius, $r_c = r_{bh} = 1/\sqrt{3} \equiv r_{max}$. For larger values of $M$, there are no positive real solutions to (\ref{roots}).

This is yet another important difference with respect to Schwarzschild black holes with $\Lambda \leq 0$, where solutions exist for all values of the mass; in dS, it is impossible to have arbitrarily large values of $M$. 

\subsection{The Nariai limit and dS$_2$}
\label{sec:nariai}
An interesting phenomenon occurs as $r_{bh}$ goes to $r_c$. This is known as the Nariai limit (the solution with $M = M_{max}$ is also known as the Nariai black hole).

Consider some value of $M$ that is close to $M_{max}$, so that the positive roots of (\ref{roots}) are $r_c = r_{max} + \varepsilon$, $r_{bh} = r_{max} - \varepsilon$, for some small $\varepsilon \ll 1$. Now we can expand the metric in between the horizons by choosing coordinates $\{\tau, \rho\}$ such that
\begin{equation}
    r = r_{max} + \varepsilon \rho \quad , \quad t = \frac{\tau}{\varepsilon} \,.
\end{equation}

\begin{figure}[h!]
        \centering
        \includegraphics[scale=0.6]{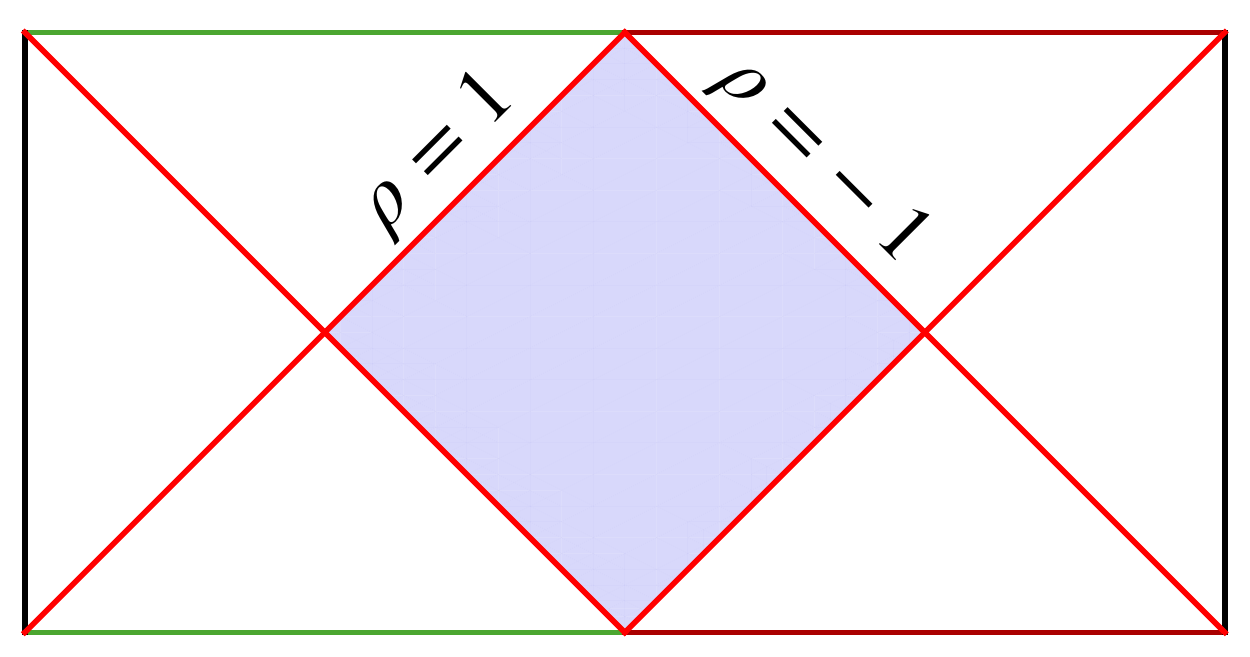}
\caption{Penrose diagram of dS in two dimensions. The shaded area can be seen as the static patch in $d=2$. The radial coordinate $\rho$ goes to $-\infty$ in the solid brown line and to $+\infty$ in the green region. This is reminiscent of the singularity and future infinity in the higher dimensional black hole. The black lines at the left and right edges are identified.}  \label{fig:nariai}
\end{figure}

To get the near horizon geometry, we expand (\ref{sds metric}) to leading order in $\varepsilon$ to obtain,
\begin{equation}
    ds^2 = - 3 (1-\rho^2) d\tau^2 + \frac{d\rho^2}{3(1-\rho^2)} + \frac{1}{3} d\Omega_{2}^2 \,,
\end{equation}
which is the metric of dS$_2 \times S^2$, with fixed radius $\ell = r_{max}$. Note that in this coordinate system, $\rho$ spans from $-1$ to $1$, where dS$_2$ has two horizons. In this sense, dS$_2$ is closer to the higher dimensional SdS geometry than to higher dimensional pure dS. In fact, the Penrose diagram of dS$_2$ looks like the SdS Penrose diagram, as can be seen in figure \ref{fig:nariai}.

It is interesting to note that the Nariai solution is actually an exact solution to the Einstein equations with a positive cosmological constant in four dimensions. Another interesting fact is that upon Wick rotating time, the topology of the solution changes with respect to the pure dS solution from $S^4 \to S^2 \times S^2$.

Finally, it is worth mentioning that this near-horizon geometry is reminiscent to the near-horizon geometry of extremal (asymptotically flat or AdS) black holes that have an AdS$_2 \times S^2$ throat. The difference is that in the dS case, we do not need charge or angular momentum for this near horizon limit to exist.

\section{Thermodynamics of de Sitter space}
Now that we know that observers in dS are surrounded by cosmological event horizons, we can study some of the properties of these horizons, pretty much in the same way as we do with black hole event horizons. This study was started in 1977 by Gibbons and Hawking in two seminal papers \cite{Gibbons:1977mu, Gibbons:1976ue}. Following those ideas, in this chapter we will find out that cosmological horizons have a temperature, an entropy obeying an area law and a particular first law of thermodynamics.

\subsection{The de Sitter temperature}
As in the case of black holes, there are many different ways to find the temperature of the cosmological horizon. It can be shown that an observer with a detector in the static patch will observe a background of thermal radiation coming from the cosmological horizon. This is nicely reviewed in \cite{Spradlin:2001pw}. Here, we stick to a purely geometrical argument in which we show that in order for the static patch geometry to be smooth in Euclidean signature, the Euclidean time needs to be periodic with a period that we identify with the inverse temperature.

Recall the metric of the static patch. Reinserting the dependence on the dS radius and taking $t \to - i t_E$, we obtain
\begin{equation}
    ds^2 = \left( 1 - \frac{r^2}{\ell^2} \right) dt_E^2 + \frac{dr^2}{\left( 1 - \frac{r^2}{\ell^2} \right)} + r^2 d\Omega_{d-2}^2 \,.
    \label{euclidean_static_patch}
\end{equation}

The idea is to look at this metric close to the horizon radius. For this, we change coordinates to $\varepsilon = \ell - r$. To leading order in $\varepsilon/\ell$, the metric (\ref{euclidean_static_patch}) becomes
\begin{equation}
    ds^2 \approx \frac{2 \varepsilon}{\ell} dt_E^2 + \frac{d\varepsilon^2}{\frac{2\varepsilon}{\ell}} + \ell^2 d\Omega_{d-2}^2 \,.
\end{equation}

Further considering the following coordinate change,
\begin{equation}
R \equiv \sqrt{2 \varepsilon \ell} \qquad , \qquad \Theta \equiv t_E/\ell \,,
\end{equation}
leaves the metric as
\begin{equation}
    ds^2 = R^2 d\Theta^2 +dR^2 + \ell^2 d\Omega_{d-2}^2 \,.
\end{equation}
The first two components of the metric look exactly like flat space in polar coordinates, but for this to be completely true, $\Theta$ needs to have the right periodicity to avoid a conical singularity at the origin. Thus, we need $\Theta \sim \Theta + 2\pi$, which leaves us with $t_E \sim t_E + 2\pi \ell$. As in statistical mechanics, we identify the period of imaginary time with the inverse temperature, so we find that
\begin{empheq}[box={\mymath[drop lifted shadow]}]{equation}
T_{dS} = \frac{1}{2\pi \ell} \,. \label{dS_temperature}
\end{empheq}

We then reach the conclusion that, as black holes, cosmological horizons also have a temperature associated to them. Note, however, that this temperature is fixed by the dS length, so it cannot be varied as in the case of black holes.

In dS, the observer cannot get rid of their horizon, so it has to be the case that the Hawking radiation of the horizon is in thermal equilibrium with its surrounding so that overall there is no evaporation. This is reminiscent of the case of eternal black holes in AdS.

Finally, considering the observed value of the cosmological constant in our own Universe, we can estimate the dS length to be around 16 billion lightyears, which will give an extremely low dS temperature of $T_{dS} \approx 10^{-30} K$.

\begin{theo}[Black hole temperature]{prf:bhtemp}
The result obtained is just a particular case of a more general family of $f(r)$ metrics that includes the asymptotically flat and asymptotically AdS Schwarzchild black holes.
Suppose we start with a metric of the form
\begin{equation}
ds^2 = -f(r) dt^2 + \frac{dr^2}{f(r)} + r^2 d\Omega_{d-2}^2 \,,
\end{equation}
where we assume that $f(r_h)=0$, and $r_h$ is the largest simple root of $f(r)$. Following an analogous procedure (Wick rotating time, expanding the metric close to the horizon and requiring smoothness of the geometry), we obtain that the temperature of the solution is given by
\begin{equation}
    T_{f(r)} = |f'(r_h)|/4\pi \,,
\end{equation}
which is also consistent with (\ref{dS_temperature}).
\end{theo}

\subsection{The de Sitter entropy} \label{sec:ds_entropy}
As with the temperature, there are different ways of computing the entropy of the cosmological horizon. In this section, we reproduce the Euclidean path integral computation, as was proposed in the second of the Gibbons-Hawking papers \cite{Gibbons:1976ue}. We will set $d=4$ and consider the Euclidean Einstein-Hilbert action with a positive cosmological constant,
\begin{equation}
    I_{E}[g] = -\frac{1}{16\pi G_N} \int d^4x \sqrt{g} \left(R-2\Lambda \right) \,,
\end{equation}
where $G_N$ is the gravitational Newton constant. The proposal of Gibbons and Hawking is to consider the following path integral
\begin{equation}
    Z = \int Dg \exp (-I_{E}[g]) \,.
\end{equation}
In principle, matter fields could be added to this path integral, but in these notes we will restrict ourselves to the purely gravitational case. See, for instance, \cite{Anninos:2020hfj, Law:2020cpj} for more general cases. The path integral, in principle, should be taken over all compact smooth geometries. As we will soon see, the round metric on the 4-sphere $S^4$ is a saddle-point solution. But we expect the path integral to be also summing over geometries with different topologies such as for instance, the Nariai solution, that has topology $S^2\times S^2$ in Euclidean signature, see section \ref{sec:nariai}.

The fact that we consider solutions with no boundaries is notorious. In fact, it is again quite different to what happens in the black hole case, where we usually consider the path integral as a function of the boundary thermal circle size $\beta$, that we interpret as the inverse temperature. See box \ref{prf:bhent} and Appendix \ref{sec:bh_entropy}.

Before continuing, we should point out that this object needs to be treated with some care as it is well-known that the gravitational path integral is somehow pathological. Gravity is a non-renormalisable theory and, among others, there is the problem of the metric conformal mode being unbounded \cite{GIBBONS1978141}. Of course these issues must be addressed in a full theory of quantum gravity.

For now, what we are going to do is to take a saddle-point approximation to the path integral, by taking $G_N \to 0$. The dominant saddle is the round metric on $S^4$. The only parameter left to fix is its radius $\ell_0$, that we need to find so as to extremise the on-shell action.

Once again, we can consider the metric (\ref{euclidean_static_patch}), for which the Ricci scalar can be computed as $R = 12/\ell^2$. Then the Euclidean action as a function of $\ell$ is given by
\begin{equation}
    I_E[\ell] = -\frac{1}{16\pi G_N} \int_0^\beta dt_E \int_0^\ell dr \int d\Omega_2  \sqrt{g} \left(R-2\Lambda \right) = -\frac{\pi}{6G_N} (12\ell^2 - 2 \Lambda \ell^4) \,.
\end{equation}
Extremising the action with respect to the radius gives
\begin{equation}
    \partial_\ell I_E[\ell] |_{\ell=\ell_0} = 0 \rightarrow 24 \ell_0 - 8 \Lambda \ell_0^3 = 0 \rightarrow \ell_0^2 = \frac{3}{\Lambda} \,.
\end{equation}
So here we recover again the result exhibited in the first chapter, see (\ref{einstein_eqs}) for $d=4$. Now plugging in the on-shell value for the cosmological constant, we find that
\begin{equation}
    I_{E}[\ell_0] = - \frac{\pi \ell_0^2}{G_N} = - \frac{A_H}{4G_N} \,,
\end{equation}
where $A_H$ is the area of the cosmological horizon. Usually, we interpret $I_E$ as $F/T$, where $F$ is the free energy of the system. But we know using thermodynamic relations that $F/T = E/T - S$, where now $E$ and $S$ are the energy and the entropy. However, as pointed out by Gibbons and Hawking, in dS there cannot be any energy, as energy is a boundary term in General Relativity and there is no boundary to define it. In this sense, it looks like we are computing a microcanonical partition function, as we are fixing the energy to be zero. And so, basically, what we computed is just minus what is called the Gibbons-Hawking entropy,
\begin{empheq}[box={\mymath[drop lifted shadow]}]{equation}
    S_{GH} = \frac{A_H}{4G_N} \,. \label{GH_entropy}
\end{empheq}

Understanding the microscopic origin (if there is one) of this formula and its quantum corrections might be one of the keys to understand holography in dS spacetime. In fact, it was probably the analogous formula for black holes which served as one of the main motivations towards what we now know as the AdS/CFT correspondence \cite{Strominger:1996sh}. A few comments about this formula:
\begin{itemize}
\item First, if we reinsert all the natural constants, the formula for the dS entropy in four dimensions becomes the one presented in (\ref{dS_entropy}),
\begin{equation}
    S_{GH} = \frac{3\pi}{\Lambda} \frac{k_B c^3}{\hbar G_N}\,,
\end{equation}
which relates the entropy of the cosmological horizon to all the fundamental constants in nature, linking this object to statistical mechanics (through Boltzmann constant $k_B$), relativity (through the speed of light $c$), quantum mechanics (through $\hbar$), gravity ($G_N$) and cosmology ($\Lambda$). The dS temperature, for instance, is similar, but it does not include gravity as it does not have $G_N$ (nor $k_B$).
    \item One of the big open questions is whether this quantity is really an entropy. If it is, according to Boltzmann, it should be counting a number of microstates. But, what is this formula counting? This is even more mysterious than in the black hole case, because the cosmological horizon is observer dependent and each observer has their own horizon with their own entropy. So what are we counting? Can it be an entanglement entropy \cite{Anninos:2020hfj}?
    \item It is huge. If we plug our current value for the cosmological constant (and the other constants in nature that we usually set to 1), we obtain that $S_{GH} \sim 10^{122} k_B$. The entropy of all the matter and energy content in our visible Universe -- that is dominated by the masses of supermassive black holes -- is estimated to be of order $S_{\text{Universe}} \sim 10^{104} k_B$ \cite{Egan_2010}. The dS entropy is still much larger, maybe incorporating the entropy of spacetime itself.
    \item Finally, it is interesting to compare it with the entropy of the dS black hole that we studied in section \ref{sec:sds}. In that case, the area is proportional to the horizon radius squared, and so it follows from figure \ref{fig:sds_roots} that it increases until reaching the largest value that corresponds to the Nariai black hole. As previously mentioned, the horizon radius of the Nariai black hole is $r_{\text{Nariai}} = \ell/\sqrt{3}$, which gives an entropy of 
    \begin{equation}
        S_{\text{Nariai}} = \frac{\pi \ell^2}{3 G_N} = S_{GH}/3 < S_{GH} \,.
    \end{equation}
    Note that even if one considers the total entropy of SdS as the sum of the areas of both the cosmological and the black hole horizons, this is still less than the entropy of empty dS. For instance, for the Nariai case, this will give $2 S_{\text{Nariai}} = 2S_{GH}/3 < S_{GH}$. So yet another interesting feature of dS is that the empty dS solution is the most entropic configuration among black hole solutions with positive $\Lambda$.
\end{itemize}

\begin{theo}[Black hole entropy]{prf:bhent}
We can also repeat the Euclidean path integral computation to obtain the entropy formula for black holes, that is also famously given by an area formula,
\begin{equation}
S_{BH} = \frac{A_H}{4G_N} \,. \label{area_law}
\end{equation}
But even if the final result is the same, in the black hole case the computation is more subtle and, in fact, the entropy comes from a boundary term in the action. Consider the Euclidean gravitational action (with no cosmological constant),
\begin{eqnarray}
    I_E = - \frac{1}{16\pi G_N}\int d^4x \sqrt{g} \, R - \frac{1}{8\pi G_N} \int_{r=r_0} d^3x \sqrt{h} \, K \,,
\end{eqnarray}
where $h$ is the induced metric at the boundary and $K$ is the trace of the extrinsic curvature.\footnote{A useful compendium of formulas to find formal definitions of all these terms can be found in \cite{compendium}.} The second term is called the Gibbons-Hawking-York (GHY) boundary term and, in manifolds with boundaries, is needed for the variational principle to be well-defined \cite{yorkbdy, Gibbons:1976ue}.

One saddle-point solution to this action is the Euclidean black hole metric,
\begin{equation}
    ds^2 = \left(1- \frac{2M}{r}\right) dt_E^2 + \frac{dr^2}{\left(1- \frac{2M}{r}\right)} + r^2 d\Omega_2^2 \quad , \quad r \in [r_h = 2M, \infty] \,. \label{sch_bh}
\end{equation}
As discussed, regularity at the horizon imposes that $t_E \sim t_E + 8\pi M$. Given that for this solution $R=0$, the bulk term in the Einstein action vanishes on-shell. Instead, the main contributions to the gravitational path integral come from a boundary term fixed at a constant $r=r_0$ slice \cite{York:1986it}. The details of the computation, that involve regularisation of the action, are shown in Appendix \ref{sec:bh_entropy}, but the final result is that the entropy is proportional to $M^2$, which gives the area law in (\ref{area_law}).
\end{theo}

\subsection{The first law for de Sitter space}

An alternative derivation of the dS entropy can be obtained from a first law for the dS horizon, and it is analogous to the derivation of the entropy from the first law of black hole mechanics.

We want to study how the area of the cosmological horizon changes as we throw some infinitesimal energy, $dM$ into the cosmological horizon. As there is no spatial boundary in dS, we need to be cautious in what we mean by energy, but at least infinitesimally, we will assume that the change in the horizon area due to dropping that energy is the same as that of having a small $M$ parameter in the Schwarzschild de Sitter black hole solution (\ref{sds metric}). More elaborate arguments can be found in Gibbons and Hawking original paper \cite{Gibbons:1977mu} and in \cite{Bousso:2002fq}. We obtain that the variation in the horizon area is given by
\begin{equation}
  \left.  d A_H \right|_{A_H = 4\pi \ell^2} = - 8\pi\ell dM \rightarrow \left. \frac{d A_H}{dM} \right|_{A_H = 4\pi \ell^2} = - \frac{4}{T_{dS}} \,,
\end{equation}
where for the last equality we used the fact that the dS temperature is given by (\ref{dS_temperature}). Assuming that the entropy is proportional to the area, we obtain a first law for the cosmological horizon,
\begin{empheq}[box={\mymath[drop lifted shadow]}]{equation}
    dM = - T_{dS} \, dS_{GH} \,, \label{first_law}
\end{empheq}
if we fix the proportionality factor to $1/4$, confirming (\ref{GH_entropy}).

Note that compared to the usual first law of thermodynamics, there is a minus sign in (\ref{first_law}). As we discussed in section \ref{sec:sds}, this is a consequence of the fact that the cosmological horizon radius decreases in size as we include some mass parameter $M$. Then, as the mass crosses the event horizon, the entropy increases, maybe signaling that the observer now has less information about the interior of the cosmological horizon \cite{Anninos:2012qw}.

As with the black hole case, it is possible to study quasi-local thermodynamics of the de Sitter horizon with respect to data at a York boundary. For the cosmological horizon, this was first studied in \cite{Hayward:1990zm, Wang:2001gt}, where in the latter it is shown that the first law (\ref{first_law}) is recovered when shrinking the size of the York boundary towards the observer's worldline. See \cite{Banihashemi:2022htw} for a recent discussion. Timelike boundaries might play an important role in defining holography in de Sitter space, so we will come back to the discussion of this problem in section \ref{sec:holo_ds}.

\section{Geodesics in de Sitter space}

In this chapter, we will study geodesics in de Sitter space. Geodesics are probably the simplest extended objects that appear in General Relativity and as such, they contain useful information about the spacetime itself.

They are also important at a semi-classical level, in the context of quantum fields in fixed curved backgrounds. In this case, the geodesic length computes the logarithm of the two-point function of heavy free massive scalar fields $\phi$. Schematically,
\begin{equation}
    \langle \phi(X) \phi(Y) \rangle = \int DP \, e^{-m \, L[P]} \approx \sum_g e^{-m \, L_g}\,, \label{geodesic_approximation}
\end{equation}
where the path integral is over all possible paths connecting the two points $X, Y$ and the last expression is the saddle point approximation as the mass $m$ goes to infinity \cite{PhysRevD.23.2850}. $L_g$ is the geodesic length between $X$ and $Y$ and if there is more than one geodesic, in principle, we should sum over them. 

Finally, to finish this introduction to geodesics, it is worth mentioning that extremal surfaces play an important role in the context of the AdS/CFT correspondence. For instance, co-dimension 2 surfaces are related to entanglement entropy through the Ryu-Takayanagi prescription and its generalisations \cite{rangamani2017holographic} and co-dimension 1 extremal volumes are conjectured to compute quantum complexity \cite{Chapman:2021jbh}. Geodesics are co-dimension 2 objects in $d=3$, and co-dimension 1 in $d=2$.

\

\noindent \textbf{Euclidean vs. Lorentzian geodesics}. In Euclidean signature, there is always a minimal path between any two points in a smooth manifold. So, there is always, at least, a minimal length geodesic. However, in general, there is no maximal path, as the infinitesimal interval, $ds^2$, is always positive. In Lorentzian signature, $ds^2$ can have either sign (or even be null), so geodesics are not minimal anymore but locally extremal curves \cite{Wald:1984rg}.

\begin{figure}[h!]
        \centering
        \includegraphics[height=4cm]{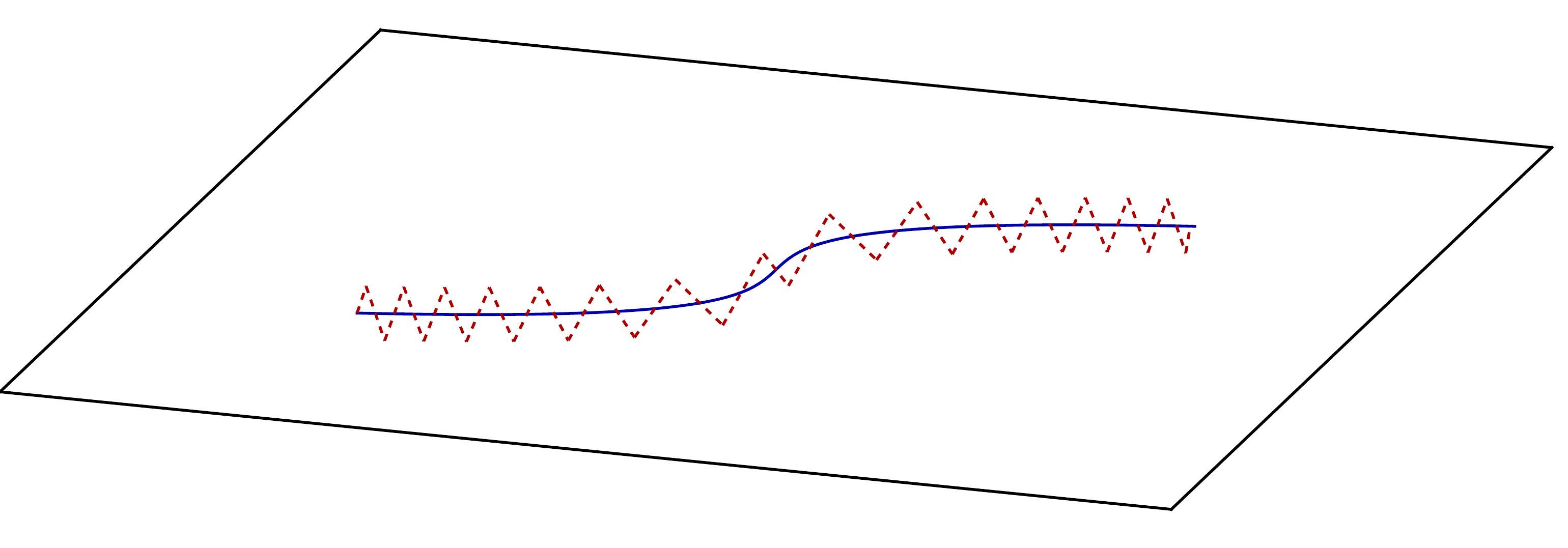}
\caption{A candidate minimal length curve between two spacelike separated points (in solid blue) and a null zig-zag curve around it that will have less length (in dashed brown). Time runs vertically in this diagram.}\label{fig:max}
\end{figure}

As an example, consider two spacelike separated points. Assume there is a geodesic between the two points that has minimal length. But now consider deforming the curve infinitesimally with a null zig-zag trajectory around the candidate geodesic, see figure \ref{fig:max}. As the zig-zag curve will have close to zero length, it will certainly have less length than the candidate geodesic and so, the candidate curve cannot be a geodesic. The formal way of saying this is that given the set of curves between two fixed points, the length (for spacelike geodesics) or the proper time (for timelike geodesics) are upper-semi continuous. This is nicely explained in Chapter 9 of \cite{Wald:1984rg}. As a consequence, in Lorentzian signature, it might be possible that certain points are not connected at all by geodesics. This happens, for instance, in dS space \cite{hawking_ellis_1973, jacobson}.

%
%
\subsection{Lorentzian geodesics in de Sitter space} \label{lor_geo}

Now we can set up our computation for geodesics in dS space. Most of this section is reviewed from \cite{Chapman:2022mqd}. For simplicity, we will consider $d=2$ and we will set $\ell=1$. We will use global coordinates (\ref{global metric}), so that the length between any two fixed points is given by the following functional,
\begin{equation}
    L = \int ds = \int d\lambda \mathcal{L} (\tau, \dot{\tau}, \varphi, \dot{\varphi}, \lambda) = \int d\lambda \sqrt{-\dot{\tau}^2 + \dot{\varphi}^2 \cosh^2 \tau } \,,
    \label{length_functional}
\end{equation}
where $\varphi$ is the angular coordinate around the spatial circle and the dot indicates derivative with respect to $\lambda$, that, for now, is some parameter along the curve. The endpoints of the curve are two arbitrary fixed points $\{\tau_0, \varphi_0\}$ and $\{\tau_1, \varphi_1\}$, so we can always choose $\lambda \in [\lambda_0,\lambda_1]$, such that
\begin{equation}
     \left\{\begin{array}{l}
        \tau(\lambda=\lambda_0) = \tau_0, \\
       \varphi(\lambda = \lambda_0 ) = \varphi_0 \,,
        \end{array}\right\} \qquad \text{and} \qquad  \left\{\begin{array}{l}
        \tau(\lambda=\lambda_1) = \tau_1, \\
       \varphi(\lambda =\lambda_1 ) = \varphi_1 \,.
        \end{array}\right\}
\end{equation}

Now the problem becomes a classical mechanics problem of extremising the length functional (\ref{length_functional}). We start by noting that $\varphi(\lambda)$ does not appear explicitly in the Lagrangian, so there is a conserved quantity $Q$ associated to it,
\begin{equation}
Q \equiv \frac{\partial \mathcal{L}}{\partial \dot{\varphi}} = \dot{\varphi} \cosh^2 \tau \left(-\dot{\tau}^2 + \dot{\varphi}^2\cosh^2 \tau\right)^{-1/2} \,. \label{conservedQ}
\end{equation}
Since the length functional is invariant under reparametrisation, we may select $\lambda$ such that it is an affine parameter, 
\begin{equation}
	\mathcal{L}^2=\left( \frac{ds}{d\lambda} \right)^2 = \pm 1 = - \dot{\tau}^2 + \dot{\varphi}^2 \cosh^2 \tau  \, , \label{energy}
\end{equation}
where the $\pm$ depends on whether the geodesic is spacelike or timelike, respectively. Note that with this convention, spacelike geodesics will have real length and timelike geodesics, imaginary. We can use (\ref{conservedQ}) to write (\ref{energy}) as the well-known equation for a particle in a potential at a fixed energy, with potential $V(\tau) = -\frac{Q^2}{2} \cosh^{-2} \tau$. It is a simple exercise to find the \textit{trajectories} for this problem, which are given by
\begin{equation}
    \tan (\varphi + \tilde{\varphi}) = \frac{Q \sinh \tau}{\sqrt{Q^2 - \cosh^2 \tau}} \,, \label{trajectory}
\end{equation}
where $\tilde{\varphi}$ and $Q$ act as integration constants that will be determined upon fixing the endpoints of the geodesic. Note that this formula assumes that $\varphi$ is a monotonic function of $\tau$. If there are turning points, $Q$ must change sign at those. Let's look at some examples.

\

\noindent \textbf{Timelike separated points.} Consider two spacetime points that are at a fixed spatial angle $\varphi_*$ but separated in time, \ie $X = \{\tau_0, \varphi_* \}$ and $Y = \{\tau_1, \varphi_*\}$. Plugging these endpoints into (\ref{trajectory}) gives $Q=0$, and a geodesic length given by $L_g = i |\tau_1 - \tau_0|$. Note that, as mentioned, timelike geodesics in this convention have imaginary lengths.

\

\noindent \textbf{Spacelike separated points.} Consider fixing the endpoints at opposite sides of the spatial circle, \ie  $X = \{\tau_0, \varphi_*\}$ and $Y = \{\tau_1, \varphi_* + \pi\}$. The only smooth solution with this choice of points has a turning point and requires that $\tau_0 = -\tau_1$. Then, there exists a one-parameter family of geodesics with $|Q|>\cosh \tau_0$. All of these geodesics have the same length, that is half the circumference of the sphere with unit radius, $L_g = \pi$. This can be also seen by noting that geodesics in the embedding picture can be obtained as intersections of the hyperboloid with a plane that contains the two points under consideration and the origin of the embedding spacetime.

\begin{figure}[h!]
        \centering
        \subfigure[dS$_2$]{
                \includegraphics[scale=0.5]{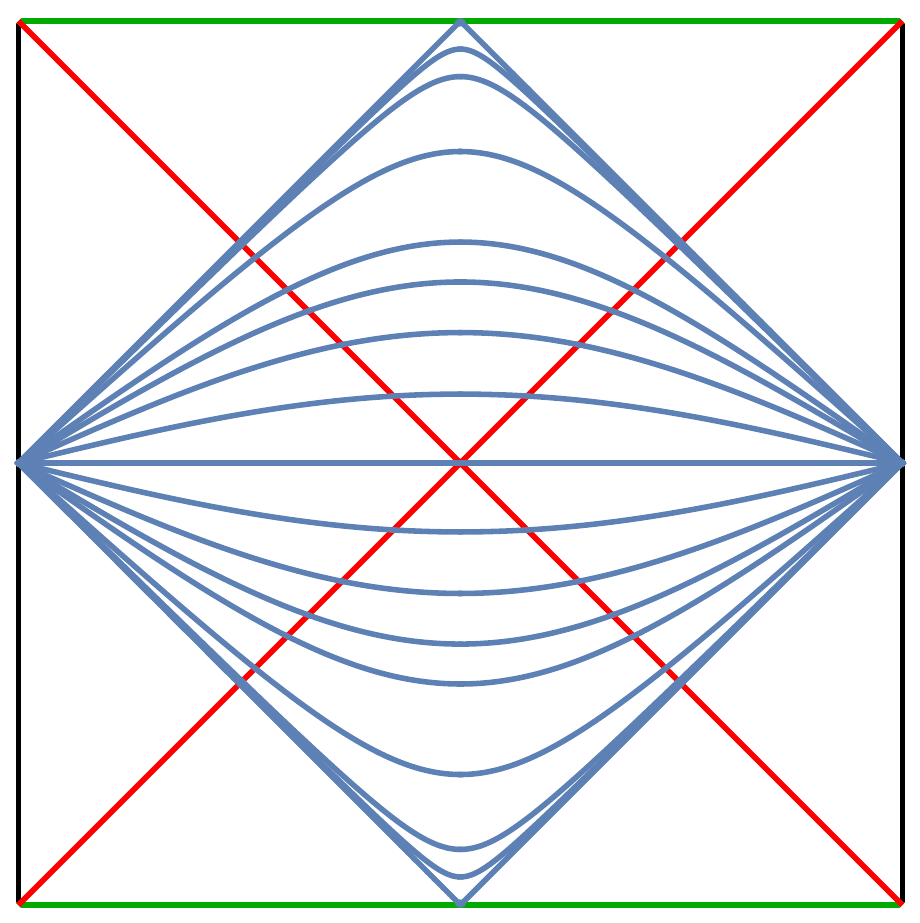} \label{fig:PenDS}} \qquad \qquad \qquad
         \subfigure[AdS$_2$ black hole]{
                \includegraphics[scale=0.5]{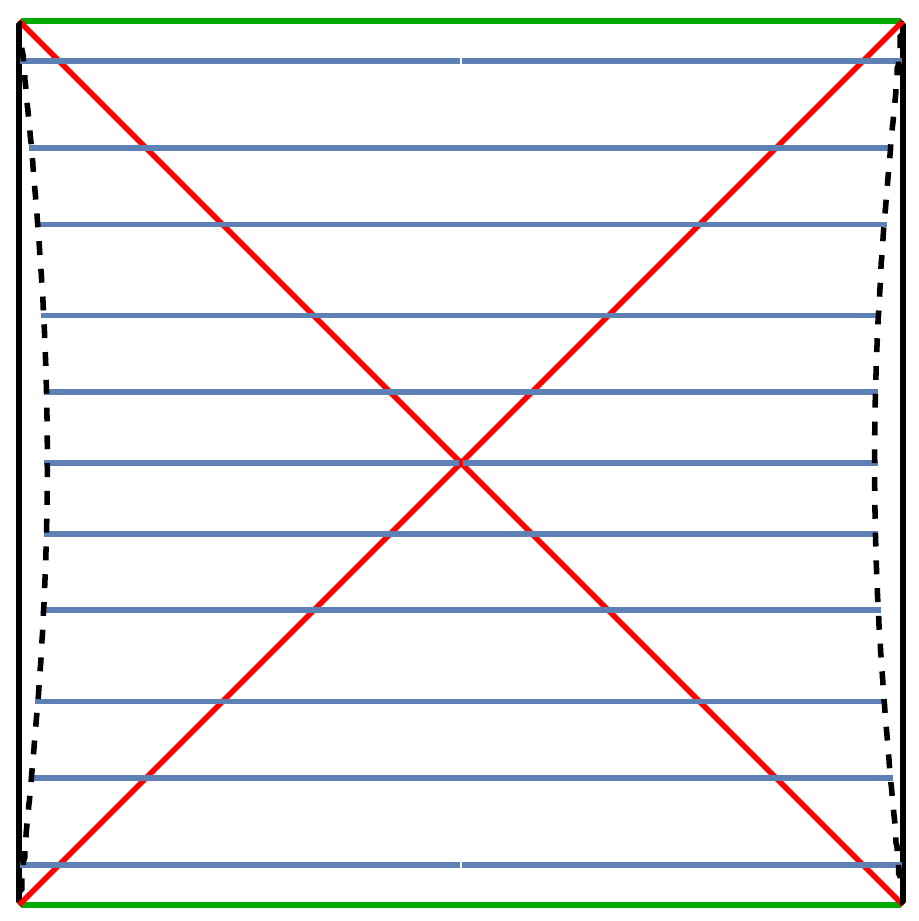}\label{fig:PenADS}} 
                 \caption{\footnotesize Penrose diagrams for (half of) dS$_2$ and the AdS$_2$ black hole. Geodesics are plotted in blue and $R_b=10$ is in dashed black in the AdS case. The red lines correspond to the horizons. Figures from \cite{Chapman:2021eyy}.}
\end{figure}

For any other choice of times, there are no real geodesics between the two points. In particular, if we consider $\tau_0 = \tau_1$, then geodesics only exist if $\tau_0 = \tau_1 = 0$. See figure \ref{fig:PenDS}.

It is worth noting that a similar feature occurs in the double-sided eternal AdS black hole in dimensions higher than 3, where real spacelike geodesics anchored at opposite boundaries stop existing at a finite time, encoding some signatures from the black hole singularity \cite{Fidkowski:2003nf, Festuccia:2005pi}.

\begin{theo}[Spacelike geodesics in AdS]{theo:theo3}
It is possible to do a completely analogous computation but now in the AdS black hole. In two dimensions, the computation is pretty straightforward, because the geodesics are those of global AdS$_2$ \cite{Brown:2018bms}. One difference with respect to the dS case is that the geodesic length diverges close to the conformal boundary, so it needs regularisation. But most importantly, if we fix the same times at each boundary, let's call it $t$, it can be shown that the geodesic length is given by
\begin{equation}
    L_g^{\text{AdS}} (t) = 2 \log \left( 2 R_b \cosh t/2 \right) + O(1/R_b^2) \approx 2 \log R_b + |t| + \cdots \,,
\end{equation}
where $R_b \gg 1$ is a large cutoff in units of the AdS scale. Note that this formula is valid for arbitrarily long times, where the length changes linearly with time, making it compatible with the holographic complexity prediction \cite{Brown:2018bms}. The geodesics are shown in figure \ref{fig:PenADS} where we contrast them with the dS ones.
\end{theo}

\

\noindent \textbf{Generic points.} The computation can be further generalised to arbitrary points in dS$_2$. Consider two points $X = \{\tau_0, \varphi_0 \}$ and $Y = \{\tau_1, \varphi_1 \}$. We do not attempt here to provide a full derivation, but it can be shown that whenever a geodesic exists, its length is given by
\begin{equation}
    L_g = \text{arccos} \left( \cosh \tau_0 \cosh \tau_1 \cos (\varphi_1 - \varphi_0) - \sinh \tau_0 \sinh \tau_1 \right) \,. \label{ds_geodesic_length}
\end{equation}
But given a particular pair of points in dS, how do we know if geodesics exist at all? It turns out that a better diagnostic of the distance between two points in de Sitter is what is inside the argument of the arccos in (\ref{ds_geodesic_length}). 

This quantity is well-defined for \textit{any} two points in dS and it can be further generalised to higher dimensions. The easiest way to discuss it is in embedding space, see chapter \ref{ch:what_is}, where it can be defined in a coordinate invariant way. We call this quantity $P_{X,Y}$, the de Sitter invariant length between two points $X, Y$ and for $d$-dimensional dS space, it is defined as
\begin{equation}
	P_{X,Y} \equiv \eta_{IJ} X^I Y^J , \quad \eta_{IJ} = \text{diag} \underbrace{(-1, 1, \ldots, 1)}_{d + 1}  \,, \label{PXY}
\end{equation} 
where $X^I$ and $Y^J$ are the coordinates in $(d+1)$-Minkowski space of any two points in the dS hyperboloid. It is clear by looking at (\ref{ds_geodesic_length}) that $P_{X,Y}$ is a real quantity that can go from $-\infty$ to $\infty$. Depending on the relative position between the two points, we have that
\begin{equation}
\begin{cases}
P_{X,Y}  >1 \, ,  &\text{for timelike separated points} \,; \\
P_{X,Y}  =1  \, ,  &\text{for coincident or null separated points} \,; \\
P_{X,Y}  <1  \, ,  &\text{for spacelike points} \,; \\
P_{X,Y}  =-1 \, ,  &\text{when $X$ is null separated from the antipodal point of $Y$} \,; \\
P_{X,Y}  <-1 \,  ,  &\text{when $X$ is timelike separated from the antipodal point of $Y$} \,.
\end{cases} \nonumber
\end{equation}
To answer the question we started with, geodesics in dS$_d$ only exist when $P_{X,Y} \geq -1$. Coming back to the example where we considered opposite points on the spatial circle at the same global time $\tau$, we find that $P_{X,Y} = - \cosh \tau$, so it becomes manifest that real geodesics only exist at $\tau=0$ (which gives $L_g = \text{arccos} (-1) = \pi$, as we found).

\subsection{Euclidean geodesics in de Sitter space}
\begin{figure}[h!]
        \centering
        \includegraphics[height=4cm]{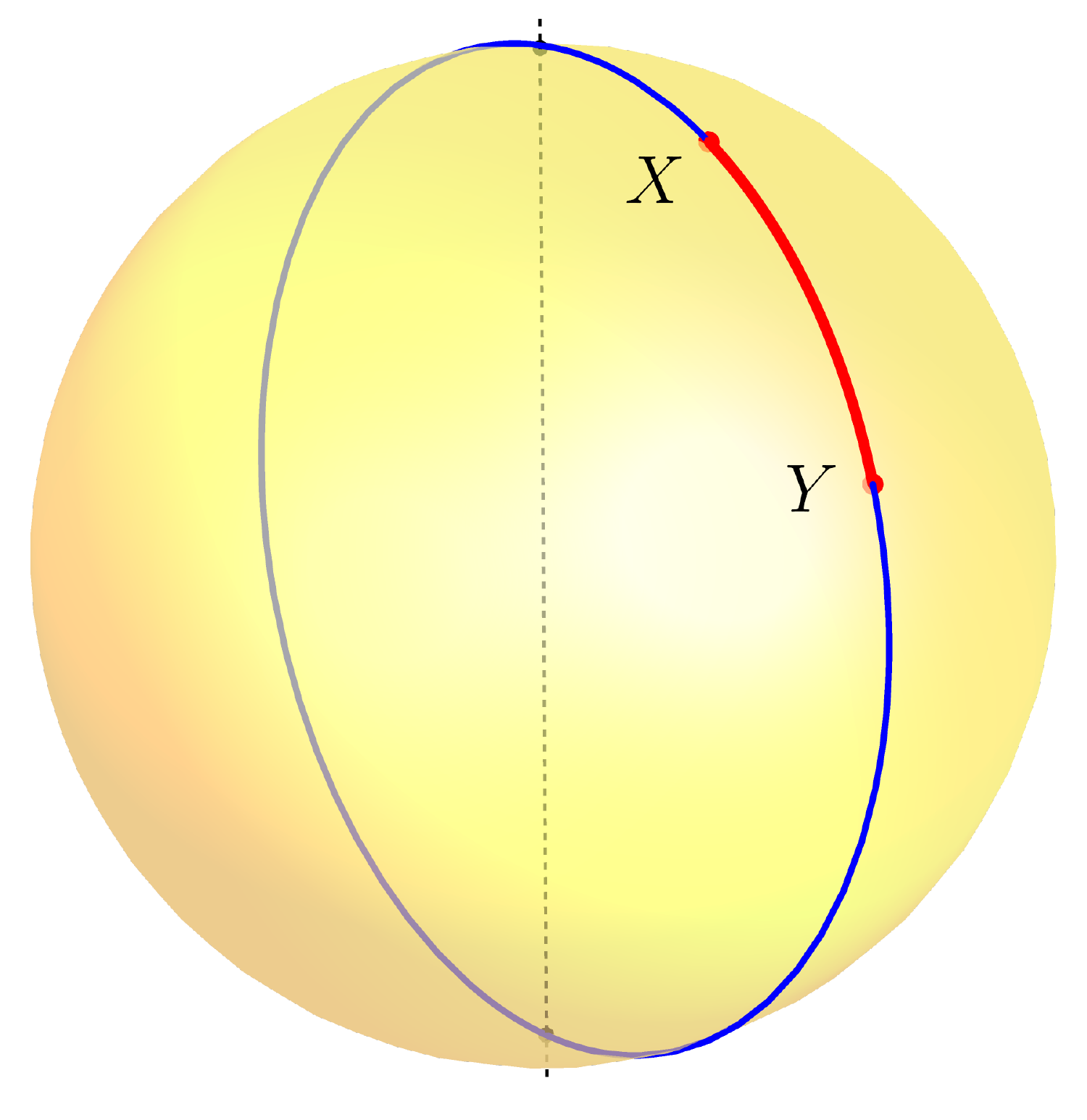}
\caption{Geodesics (in red and blue) between two points $X$ and $Y$ in the sphere form a great circle.}\label{fig:geo_sphere}
\end{figure}

As previously discussed, in Euclidean signature there is no issue with finding geodesics between any two points. In section \ref{sec:sphere}, we discussed that the analytic continuation of the time coordinate maps dS$_d$ space into the sphere, $S^d$. Here, the problem of finding geodesics is famously solved, \ie between any two points there exists a great circle that connects them. Both segments connecting the two points along the great circle are geodesics and the one with minimal length is the shortest path between the two points. See figure \ref{fig:geo_sphere}. There are also infinitely many other geodesics that wrap around the great circle.

We can now define our invariant distance in Euclidean space, $P_{X,Y}^E = \delta_{IJ} X^I Y^J$, that is simply related to the geodesic length by $P_{X,Y}^E = \cos L_g$. In two dimensions, we can use the coordinates in (\ref{global_sphere}), taking the angular coordinate $\varphi$ to parameterise the spatial circle, to show that,
\begin{equation}
    P_{X,Y}^E = \cos \tau_{E,0} \cos \tau_{E,1} \cos (\varphi_1 - \varphi_0) - \sin \tau_{E,0} \sin \tau_{E,1} \,.
\end{equation}
Note that $-1 \leq P_{X,Y}^E \leq 1$, so the geodesic length $L_g$ is always well-defined, and that upon analytically continuing back to Lorentzian time, $P_{X,Y}^E \to P_{X,Y}$. This fact (and other subtleties of the analytic continuation) was recently used in \cite{Chapman:2022mqd} to understand the geodesic approximation (\ref{geodesic_approximation}) in the Lorentzian cases where no real geodesics exist.

\section{Shockwaves in de Sitter space} \label{chap:shocks}

So far we have studied different solutions -- pure dS, Schwarzschild dS, Nariai -- to the Einstein equations with a positive cosmological constant and no matter content. In this chapter, we will study the (possibly) simplest solutions with a non-vanishing stress tensor, and we will explore their consequences as a particular case of what is known as Gao-Wald theorem \cite{Gao:2000ga}.

Consider solutions to Einstein equations with a positive cosmological constant sourced by some matter stress tensor $T_{\mu\nu}$, that we will specify shortly,
\begin{equation}
    R_{\mu\nu} - \frac{1}{2} R g_{\mu\nu} + \Lambda g_{\mu\nu} = 8 \pi G_N T_{\mu\nu} \,. \label{ee_stress}
\end{equation}

For now, we can work in any spacetime dimensions. We are interested in massless sources that move along null directions, so it will be useful to consider null coordinates like the ones presented in (\ref{kruskal_coords}). For completeness, we remind ourselves that in the absence of matter, the pure dS solution in these coordinates is given by
\begin{equation}
    ds^2 = \frac{-4 dUdV + (1+UV)^2 d\Omega_{d-2}^2}{(1- UV)^2} \,.
\end{equation}

We want to study how the geometry changes when we insert matter in the form of a shockwave. The shockwave stress tensor that we will consider simply takes the form of a delta function in one of the null directions,
\begin{equation}
    T_{\mu\nu} = \frac{(d-2)}{4\pi G_N} \alpha \, \delta(U) \delta^U_\mu \delta^U_\nu \,. \label{shock_stress}
\end{equation}
To satisfy the null energy condition, we require $\alpha > 0$. Note that this shock is localised in the $U-$direction, but not in the compact directions, so it is basically a spherically symmetric shell of null matter with the size of the cosmological horizon. Given the divergence of the stress tensor at $U=0$, one might find it surprising that there exist analytic solutions sourced by a delta function stress tensor. However, shockwave solutions have been known for some time. In asymptotically flat spacetimes, exact shockwave solutions were first discussed in \cite{sexl, DRAY1985173}. The generalisation to the case with positive cosmological constant was discussed first by Hotta and Tanaka \cite{Hotta:1992qy, Hotta:1992wb}, and Sfetsos. \cite{Sfetsos:1994xa}.

Shockwave solutions can be obtained by noting that away from $U=0$, the solution has to be a solution of the vacuum Einstein equations. For now, we will consider the case where both for $U>0$ and $U<0$, the spacetime is pure de Sitter. The matching should be such that the Einstein equations are satisfied at the junction \cite{Penrose:1972xrn}. The solution after the shock looks like a shift in the $V$-coordinate,
\begin{equation}
    V_{U>0} = V_{U<0} - \alpha \,.
\end{equation}
Using the metric in (\ref{kruskal_coords}), it is possible to show that the shockwave metric is given by,
\begin{equation}
     ds^2 = \frac{-4 dUdV + (1+U(V-\alpha \theta(U))^2 d\Omega_{d-2}^2}{(1- U(V-\alpha \theta(U))^2} \,,
\end{equation}
where $\theta$ is the Heaviside function. This is called the Rosen form of the metric. Sometimes it is more convenient to use discontinuous coordinates $u = U$, $v = V - \alpha \theta(U)$, where the metric takes the more standard shockwave form,
\begin{equation}
   ds^2 = -\frac{4}{(1- uv)^2} dudv - 4 \alpha \delta(u) du^2 + \frac{(1+uv)^2}{(1- uv)^2} d\Omega_{d-2}^2 \,. \label{ds_shock}
\end{equation}
It is easy to check that this metric satisfies the Einstein equations sourced by stress tensor (\ref{shock_stress}).

\begin{figure}[h!]
        \centering
        \subfigure[Shockwave in dS]{
                \includegraphics[height=4cm]{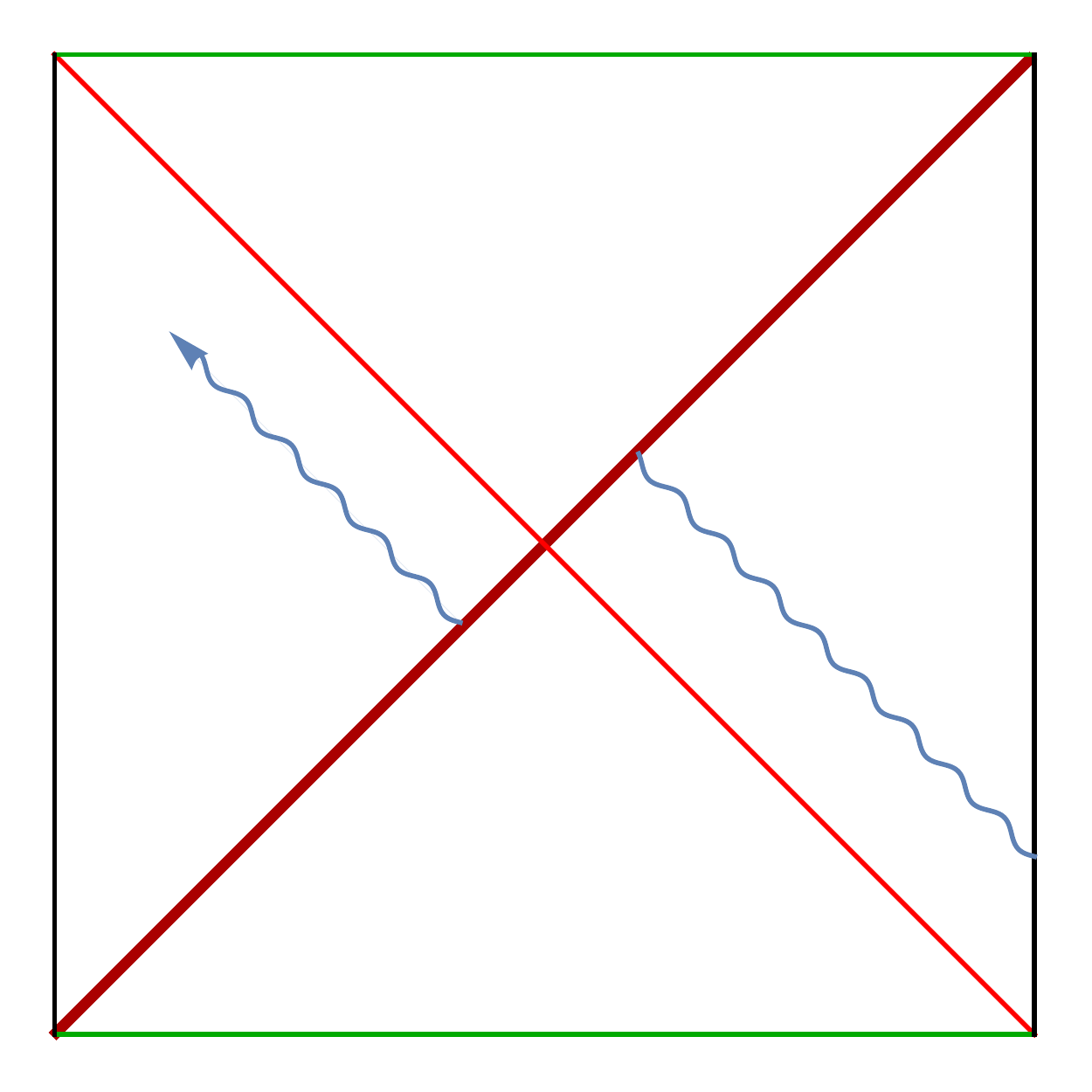} \label{fig:X1}} \quad 
         \subfigure[Shockwave in dS]{
                \includegraphics[height=4cm]{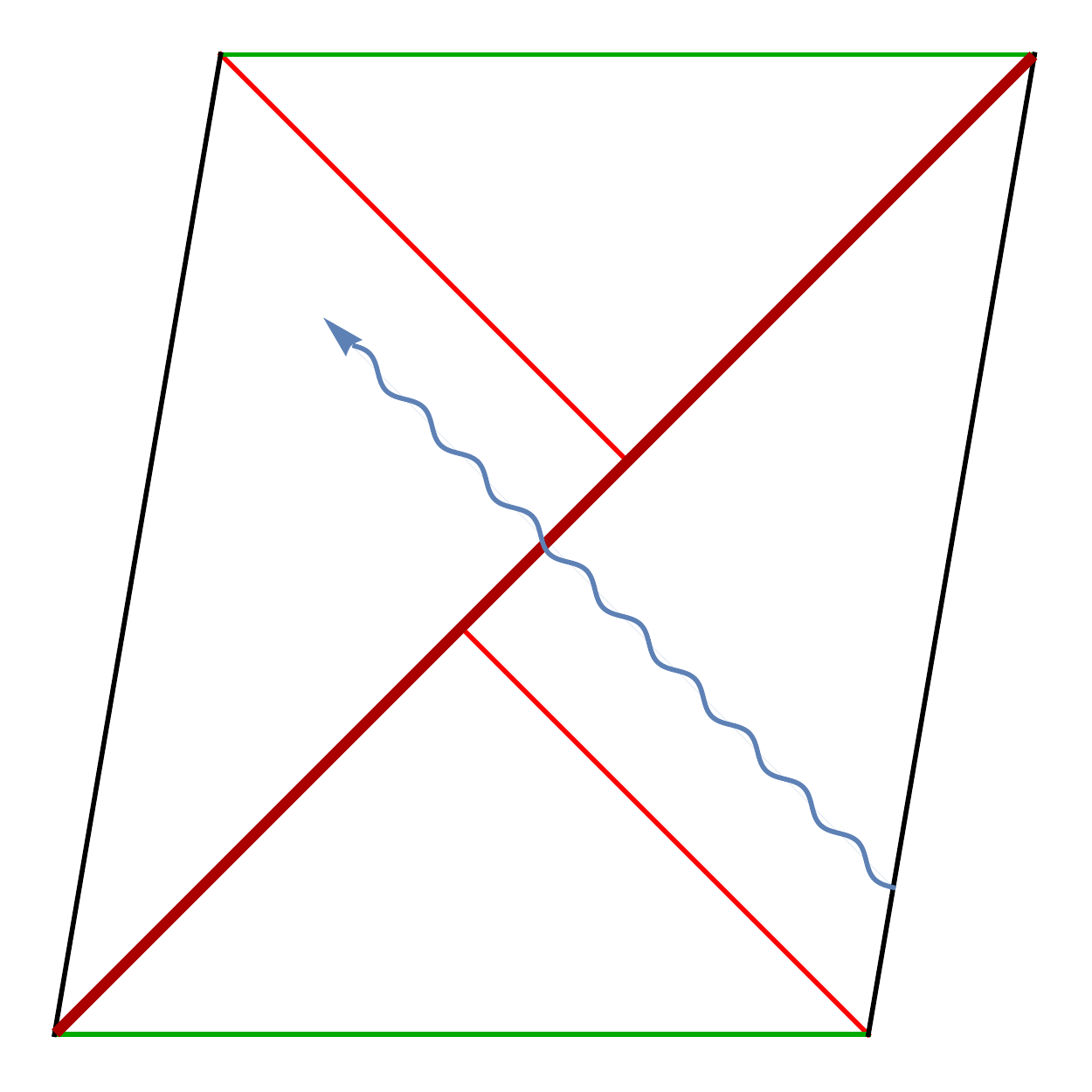}\label{fig:PenX2}} \quad 
         \subfigure[Shockwave in AdS]{
                \includegraphics[height=4cm]{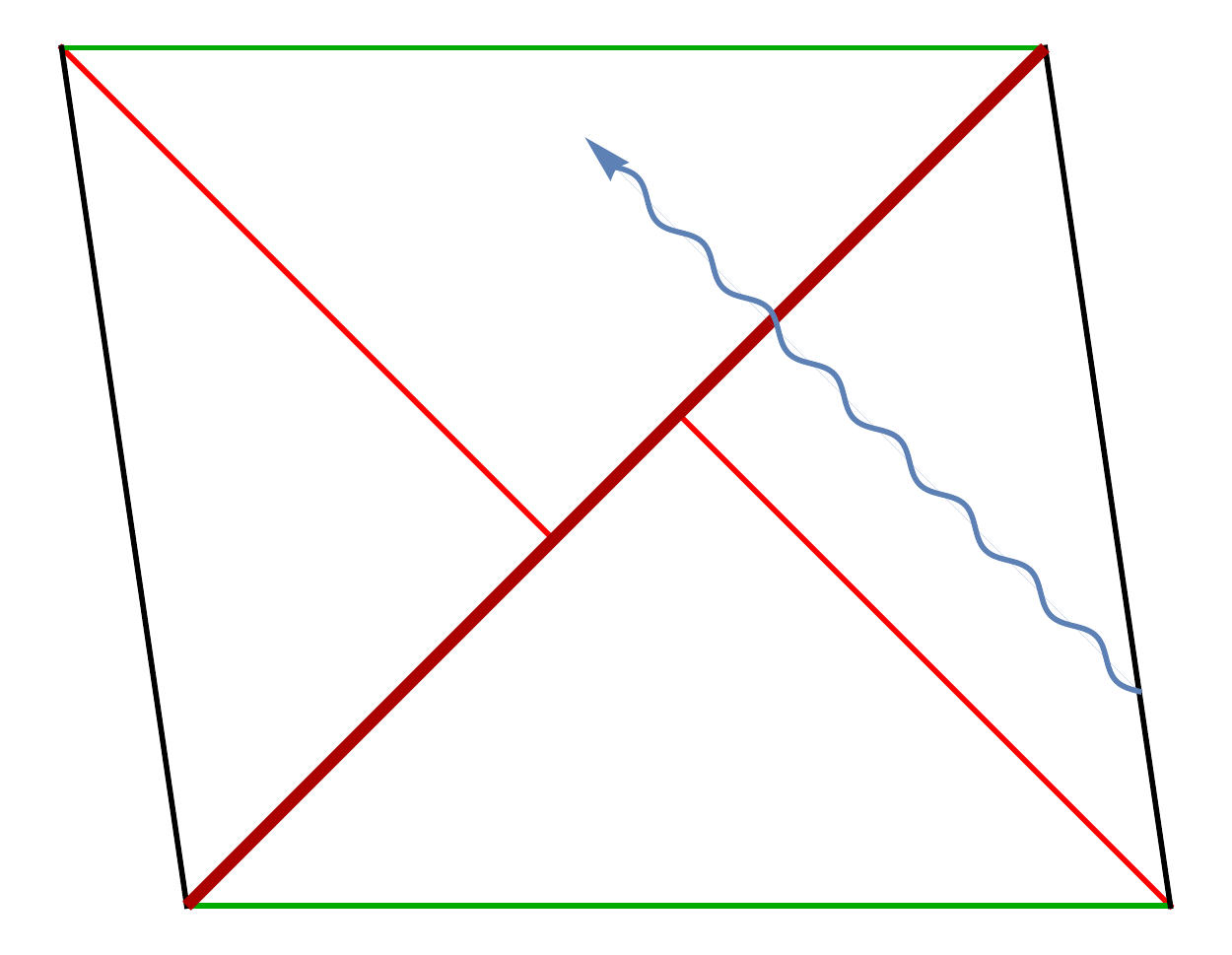}\label{fig:ads_shock}} 
                 \caption{\footnotesize Penrose diagrams for (positive energy) shockwave solutions in (A)dS. The shockwave is depicted as a dark red, thicker line. The first two diagrams depict the dS shockwave in the two different ways discussed. In (c), we plot the shockwave solution in AdS. We can see that while the dS diagram gets taller, the AdS one becomes wider.} \label{fig:shocks}
\end{figure}

Note that the sign in the term proportional to $\alpha$ in (\ref{ds_shock}) is negative, as opposed to the case of flat or AdS shockwaves -- see box \ref{theo:ads_shocks} for the AdS solution. Because of this, a light particle passing near the shock will experience a Schapiro time advance, making it possible to access a region that was causally inaccessible before. See figure \ref{fig:shocks}. This is again in clear contrast with shockwave solutions in flat or AdS spacetimes, where the particle would experience a time delay and will become even more disconnected from the other side. In fact, in those cases, the only way to make this wormhole traversable is by violating the null energy condition, \ie setting $\alpha<0$. An example of this kind in AdS is the so-called Gao-Jafferis-Wall wormhole \cite{Gao:2016bin}. 

The solution (\ref{ds_shock}) has some interesting features. For instance, the solution is pure dS space (with the same dS length) on both sides of the shock. In between there is a positive energy, spherical shock with the same size as the dS horizon. This exists at the horizon from $\mathcal{I}^-$ all the way to $\mathcal{I}^+$.

It would be interesting to find other shockwave solutions in dS. For instance, with shocks sent at a finite time. It would also be desirable to have dS analogues to the multiple shocks solutions in AdS \cite{Shenker:2013yza}, or more localised shockwaves. Lower dimensional spacetimes might provide an interesting avenue to explore these generalisations. 

\begin{theo}[Shockwaves in AdS black holes]{theo:ads_shocks}
A very similar computation can be done for black holes in AdS$_3$ \cite{Shenker:2013pqa}. The solution to (\ref{ee_stress}) with stress tensor (\ref{shock_stress}) but now with \textit{negative} cosmological constant is given by
\begin{equation}
   ds^2 = - \frac{4}{(1+ uv)^2} dudv + 4 \alpha \delta(u) du^2 + \frac{(1-uv)^2}{(1 + uv)^2} d\varphi^2 \,. \label{ads_shock}
\end{equation}
In this coordinate system, the AdS boundaries are at $uv = -1$ and the singularities at $uv=1$. Note that the sign in the term with $\alpha$ has flipped, generating the aforementioned Schapiro time delay. 

We can also plot the Penrose diagram in the AdS black hole case that, as shown in figure \ref{fig:ads_shock}, does not become taller as in the dS case, but wider, making it even harder for disconnected observers to communicate.
\end{theo}

Finally, it is worth mentioning that the effect presented is not particular to shockwaves, but a more general statement about perturbations of dS. In fact, it is proven that, in dS, all perturbations obeying the null energy condition and the null generic condition, will make the dS Penrose diagram \textit{taller}, allowing for previously disconnected regions of spacetime to become causally connected. See, for instance, \cite{Leblond:2002ns}. The formal statement goes under the name of \textbf{Gao-Wald theorem} \cite{Gao:2000ga} and states that, for solutions obeying the null energy condition and null generic energy condition, a null geodesically complete and globally hyperbolic spacetime with compact Cauchy surfaces cannot exhibit a particle horizon. In this case, sufficiently late-time observers will see the full Cauchy slice in their past light cone.

Regarding holography, in the case of asymptotically AdS spaces, shockwave solutions have played an important role in diagnosing the quantum chaotic nature of gravity. In fact, they are an essential component in the computation of out-of-time-ordered correlators that led to maximal chaos in AdS black holes \cite{Shenker:2013pqa}. Given that shockwaves in dS  behave in a different (almost opposite) way, it would be interesting to understand the consequences of this for a putative holographic theory for dS. We will discuss more about this in chapter \ref{sec:holography}.

\section{Towards holography in de Sitter space} \label{sec:holography}
Having described some of the main (semi-)classical features of dS space, in this section we will review some recent advances made in understanding quantum features of dS space. As already noted, there is still not a unified framework as in the case of the AdS/CFT correspondence, but in this Chapter we discuss some of the recent attempts to define holography in dS.

We start with a short discussion in section \ref{sec:QFT} about quantum field theory in a fixed dS background, which has gained renewed attention recently. When it comes to holography, the first question that appears is how to think about the boundary theory. In dS, the only asymptotic boundaries are at $\mathcal{I}^{\pm}$, that are spacelike. The initial proposals for holography in dS considered a Euclidean CFT living at those boundaries. This is the idea behind what is called the dS/CFT correspondence, which we also briefly review in \ref{sec:QFT}.

Most of the discussion in this Chapter will be centered around new developments towards a more local version of dS holography, which we discuss in section \ref{sec:holo_ds}. We discuss some static and dynamical features of the cosmological horizon and compare them with results in holographic studies of black holes. In this observer's approach to holography in dS, it seems that timelike boundaries might play an important role, so we finish this chapter by discussing some relevant aspects of timelike boundaries that depend on the number of spacetime dimensions under consideration.

\subsection{QFT in de Sitter space and the dS/CFT correspondence} \label{sec:QFT}
A first step towards a full quantum gravity theory in dS is to study quantum field theory in a fixed dS background. Given that in dS there is no global timelike Killing vector, even the notation of a vacuum state needs to be considered carefully, as, for instance, particle number is not conserved. 

However, there is in dS one particular state that is called the \textbf{Bunch-Davies} or \textbf{Euclidean} state that has very interesting properties \cite{Bunch:1978yq}. This state is invariant under the dS isometries and correlation functions exhibit no singularities except at coincident (or null separated) points, where the behaviour mimics the singularities in flat space. Moreover, it is the state that can be obtained from analytic continuation from the sphere, which is the reason behind its name.

As the simplest example, we will briefly study free massive scalar field theory on a fixed dS background. The action for this theory is given by
\begin{eqnarray}
    I_\phi = \frac{1}{2} \int d^d x \sqrt{-g} \left( g^{\mu\nu} \partial_\mu \phi \partial_\nu \phi + m^2 \phi^2 \right)\,,
\end{eqnarray}
where $m$ is the mass of the scalar field $\phi$ and $g_{\mu\nu}$ is the $d$-dimensional dS metric. This theory has been extensively reviewed in, for instance, \cite{Spradlin:2001pw, Anninos:2012qw}. In the Euclidean vacuum state $|E\rangle$, the Wightman two-point correlator is known analytically and is given by
\begin{equation}
    \langle E | \phi(X) \phi(Y) | E \rangle = \frac{\Gamma (\Delta) \Gamma (\bar{\Delta})}{(4\pi)^{d/2} \Gamma\left(d/2\right)} \, _2F_1 \left( \Delta, \bar{\Delta}; \frac{d}{2}; \frac{1+P_{X,Y}}{2} \right) \,,
\end{equation}
where $_2F_1$ is the hypergeometric function and the correlator only depends on $P_{X,Y}$, that is the dS invariant length, see section \ref{lor_geo}. It is clear, then, that this correlator is invariant under the dS isometries. The $\Delta$'s are called scaling dimensions and are given in terms of the mass and the spacetime dimension as
\begin{equation}
    \Delta = \left( \frac{d-1}{2} \right) + \sqrt{\left( \frac{d-1}{2}\right)^2 - m^2} \quad , \quad \bar{\Delta} = d-\Delta \,, \label{deltas}
\end{equation}
where $\bar{\Delta}$ is the shadow dimension. Remember here the mass is given in units of the dS radius $\ell$. These are very similar to the AdS scaling dimensions, but differ on the minus sign inside the root, which, in AdS, makes all the $\Delta \geq 0$. On the contrary, in dS, we can distinguish, in principle, two types of fields: light fields with $m < (d-1)/2$ have real $\Delta$, while heavy fields with $m > (d-1)/2 $ have complex scaling dimensions.

These complex scaling dimensions were first thought of as signals of non-unitarity in dS. However, we know they are part of the \textbf{Unitary Irreducible Representations} (UIRs) of the dS group $SO(d,1)$. These have been studied a long time ago \cite{doi:10.1098/rspa.1947.0047}, but in the last few years, there has been a renewed interest in the subject, see \cite{Sengor:2019mbz, Hogervorst:2021uvp, Pethybridge:2021rwf, Sengor:2022lyv, Letsios:2022slc, Penedones:2023uqc, Letsios:2023qzq} for general $d$ and \cite{Kitaev:2017hnr, Anninos:2019oka, Anous:2020nxu} for $d=2$. We will not go into details of these in the present lecture notes, but refer the interested reader to \cite{Sun:2021thf} for a pedagogical introduction to the subject. In general, the UIRs depend on the spacetime dimension, $\Delta$, and the spin.  As an example, the spinless representations of the dS group in $d=4$ are as follows:\footnote{Here $\Delta$ is a label for the eigenvalue of the quadratic Casimir of the group, that in the spinless case is given by $\Delta (\Delta-(d-1))$.}
\begin{itemize}
    \item \textbf{Principal series}: $\Delta = 3/2 + i \nu$, with $\nu \in \mathbb{R}$, corresponding to heavy fields in dS.
    \item \textbf{Complementary series}: $0 < \Delta < 3$, corresponding to light fields in dS.
    \item There are also \textbf{exceptional series} for integer values of $\Delta$, that in $d=4$ coincide with the \textbf{discrete series}. The interpretation of these series in terms of particles in dS is not completely understood.
\end{itemize}

Recently, UIRs have been used as a powerful tool to compute cosmological correlators at late times in what is known as the cosmological bootstrap programme. The aim of this programme is to construct cosmological correlators using physical principles such as locality, unitarity and symmetries. See \cite{Baumann:2022jpr} for recent developments and references.

\

So far we have only considered QFT on a fixed dS background. A full holographic description of dS has to allow for spacetime fluctuations. As soon as the AdS/CFT correspondence was formulated, ideas on how to incorporate cosmology into that framework appeared. See, for instance, \cite{Fischler:1998st,Balasubramanian:2001nb}. Eventually, they converged into what is now known as the \textbf{dS/CFT correspondence} \cite{Witten:2001kn, Strominger:2001pn, Maldacena:2002vr}.

The basic idea is that in analogy to the AdS/CFT correspondence, there is a dual (Euclidean) conformal field theory that lives in the asymptotic future of dS. In this case, the partition function of the boundary CFT is dual to the wavefunction of the Universe $\Psi_{dS}$,
\begin{equation}
    \Psi_{dS} [ \varphi_0 ] \leftrightarrow Z_{CFT} [\varphi_0 ] \,,
\end{equation}
where $\varphi_0$ is some asymptotic profile for bulk fields $\varphi$, that include the metric in the Hartle-Hawking state \cite{Hartle:1983ai}. We refer the reader to interesting progress in \cite{Hartle:2007gi, Hartle:2008ng, Anninos:2012ft, Pimentel:2013gza, Anninos:2014lwa, Arkani-Hamed:2017fdk, Chakraborty:2023los}.

A concrete model for this conjecture was found in \cite{Anninos:2011ui}, where a particular higher spin theory on dS was conjectured to be dual to a \textit{simple} conformal field theory that is just a theory of $N$ free scalar fields that transform as a $Sp(N)$ vector. The tower of higher spin fields in the bulk are then dual to the higher spin conserved currents of the boundary theory. Generalisations of this model can be found in \cite{Chang:2013afa, Anninos:2014hia}. The analogous version in two-dimensional JT gravity has been explored in \cite{Maldacena:2019cbz, Cotler:2019nbi}, while similar ideas in three dimensions are discussed in \cite{Hikida:2021ese,Hikida:2022ltr}. Another set of interesting ideas relating cosmology to the standard AdS/CFT correspondence by using RG flows and analyticity followed from work initiated in \cite{McFadden:2009fg, McFadden:2010na}.

\subsection{Static patch holography and timelike boundaries} \label{sec:holo_ds}
One drawback of the global approach is that it is harder to probe the cosmological horizon and its features (some of which we discussed in previous chapters). Given the success of the holographic framework in describing black hole horizons, we might hope that some of the tools used in that case might serve useful to characterise the cosmological horizon. Using holography, we came to the idea that black holes are highly-entropic, dissipative, strongly-coupled, maximally chaotic states in their quantum description. It would be desirable to understand which of these properties (if any) are still present in the cosmological horizon case.

For this, it is necessary to focus our holographic constructions inside the static patch region of dS. This might be, in principle, problematic since there is no asymptotic boundary inside the static patch. Then, the most natural places to put the holographic theory are either the observer's worldline or a stretched horizon, a timelike surface of constant radius close to the cosmological horizon. We will call these two approaches \textbf{worldline holography} and \textbf{stretched horizon holography}, respectively.

In any case, at the static level, the first important feature to understand is the quantum origin of the Gibbons-Hawking entropy for dS. Putting together the finite dS entropy and the fact that black holes cannot have arbitrarily large mass in dS, see section \ref{sec:sds}, it was suggested that the Hilbert space of a quantum dS theory should be finite dimensional \cite{Bousso:2000nf, Bousso:2000md, Banks:2000fe, Banks:2001yp, Parikh:2004wh, Banks:2006rx, Banks:2018ypk, Susskind:2021omt, Susskind:2021dfc}. In part to better understand this proposal, corrections to the Gibbons-Hawking entropy have been computed using Euclidean path integral methods. For example, it was shown \cite{Anninos:2020hfj, Law:2020cpj} that in pure gravity in $d=4$,
\begin{equation}
    \log Z = S_{GH} - 5 \log S_{GH} - \frac{571}{90} \log \frac{\ell}{\ell_{\text{ref}}} + \mathcal{O}(1) \,.
\end{equation}
The leading term is, of course, the Gibbons-Hawking entropy. Then there are two types of logarithmic corrections. The difference between them is that the second term needs an effective theory scale $\ell_{\text{ref}}$. Some of these contributions can be interpreted as entanglement entropy for gravitational edge modes \cite{Benedetti:2019uej, David:2022jfd}. Similar expressions have been studied in $d=3$, where there exists an all-loop calculation \cite{Castro:2011xb, Anninos:2020hfj, Anninos:2021ihe} and in $d=2$, where a non-perturbative formula in timelike Liouville theory has been conjectured \cite{Polchinski:1989fn, Anninos:2021ene, Muhlmann:2021clm, Mahajan:2021nsd, Anninos:2021eit, Muhlmann:2022duj,Giribet:2022cvw}.

The observation that subleading terms in the Gibbons-Hawking entropy look like entanglement entropy suggests that it might be interesting to consider entanglement entropy in dS holographically \cite{Maldacena:2012xp}, extending the Ryu-Takayanagi prescription (and its generalisations) \cite{rangamani2017holographic}. It was noted that the bifurcation surface in dS is a minimax surface, instead of a maximin (as in AdS black holes) \cite{Susskind:2021esx, Shaghoulian:2021cef, Shaghoulian:2022fop}. This suggests that in order to get similar results to those obtained in standard AdS holography, one should anchor extremal surfaces at the (stretched) horizon. See other recent proposals in \cite{Narayan:2015vda, Arias:2019pzy, Narayan:2022afv, Doi:2022iyj,Kawamoto:2023nki}. Quantum extremal surfaces in the context of dS have also been actively studied recently \cite{Chen:2020tes,Hartman:2020khs,Balasubramanian:2020xqf,Sybesma:2020fxg,Geng:2021wcq,Kames-King:2021etp,Levine:2022wos,Aalsma:2022swk}, but it is fair to say that compared to the black hole case (see \cite{Bousso:2022ntt} for current state of affairs), the case of the cosmological horizon is still pretty much under development.

A next step to characterise the horizon is to move slightly away from equilibrium. A natural probe is to consider quasinormal mode behaviour of the cosmological horizon. The scalar quasinormal frequencies are given by \cite{Abdalla:2002hg, Lopez-Ortega:2006aal},
\begin{equation}
  i \omega_n \ell = \begin{cases}
  (l + 2n + \Delta) \\
  (l + 2n + \bar{\Delta}) \end{cases} \,, \, n \in \mathbb{N}_0 \,,
\end{equation}
where $\Delta, \bar{\Delta}$ are given in (\ref{deltas}). An interesting observation is that as the mass of the perturbation increases, then it is the real part of the quasinormal frequency that increases. This is in contrast with the black hole case, where an increase in the mass translates into a more dissipative frequency with a larger negative imaginary part.

Other out of equilibrium considerations, such as questions of quantum chaos and scrambling in dS have also been recently under debate. Historically, the existence of a horizon in de Sitter space (and the associated Rindler behaviour close to it) motivated the conjecture that quantum dS is a fast scrambler \cite{Susskind:2011ap, Geng:2020kxh, Blommaert:2020tht}, in the sense that the scrambling time $t_*$ would scale as $t_* \sim \beta \log S_{GH}$, as happens in black holes.

However, this does not seem to be the prevailing interpretation these days. In quantum systems with a large number $N$ of degrees of freedom, one diagnostic of quantum chaos is the four-point out-of-time-ordered correlator ($OTOC$), that in chaotic systems schematically behaves as
\begin{equation}
    \langle OTOC \rangle_\beta (t) = f_1 - \frac{f_2}{N} \exp \lambda_L t + \cdots \,,
\end{equation}
where $f_{1,2}$ are order one positive numbers and $\lambda_L$ is known as the Lyapunov exponent. A bound on quantum chaos implies that $\lambda_L \leq 2\pi/\beta$ \cite{Maldacena:2015waa}. A shockwave computation in an AdS black hole background shows that AdS black holes saturate this bound \cite{Shenker:2013pqa}. The analogous shockwave computation in dS gives $f_2 <0$, due to the Gao-Wald effect \cite{Aalsma:2020aib}. Moreover, the rapid approach of spacelike geodesics to future and past infinities in dS, see section \ref{lor_geo}, has led to the conjecture that actually the scrambling time in dS reduces to zero, so it has been called in \cite{Susskind:2021esx} a hyperfast scrambler. There, it was also suggested that the double scaled Sachdev-Ye-Kitaev model \cite{Berkooz:2018jqr} might have this feature. Several works have tried to make this relation more precise \cite{Susskind:2022dfz, Lin:2022nss, Lin:2022rbf, Susskind:2022bia, Rahman:2022jsf, Bhattacharjee:2022ave, Goel:2023svz, Rabinovici:2023yex, Blommaert:2023opb}; however, at the time of writing, the status of chaos and complexity in dS remains inconclusive \cite{Reynolds:2017lwq, Chapman:2021eyy, Jorstad:2022mls, Galante:2022nhj, Auzzi:2023qbm, Anegawa:2023wrk, Anegawa:2023dad, Baiguera:2023tpt, Aguilar-Gutierrez:2023zqm}. A holographic computation of the $OTOC$ in a geometry with a cosmological horizon has been done in two dimensions and gives a correlator that does not grow exponentially but oscillates in time \cite{Anninos:2018svg}, pointing toward a more organised nature of the quantum theory. 

Having at least simple toy models of quantum theories for dS might help us understand these macroscopic puzzles. So what do we know about the quantum theories?

\subsubsection{The role of the observer}

The role played by the observer in holography of the dS static patch has been emphasised in \cite{Anninos:2011af}, where it was shown that correlators along the observer's worldline were controlled by $SL(2,\mathbb{R})$ symmetries. This is compatible with the idea that there is a holographic large $N$, conformal worldline quantum mechanical dual. 

From the point of view of algebraic quantum field theory, it has recently been noted that, in the presence of gravity, it might be convenient to describe a theory in terms of the algebra of observables along the timelike worldline of an observer. Moreover, in a gravitational system that is closed, like dS, it has been shown that it is essential to explicitly incorporate the observer in the analysis to obtain well-defined notions of entropy \cite{Chandrasekaran:2022cip, Witten:2023qsv}.

Incorporating the observer makes it possible to define a trace (from which we can define entropies) and in dS, it generates a von Neumann algebra of Type II$_1$ (at least in the limit of $G_N \to 0$). We will not discuss von Neumann algebras in these lectures, but we refer the reader to \cite{Sorce:2023fdx} for a pedagogical introduction. Instead, we will just point out a few properties that might resonate with some ideas already discussed in these notes.

This type of algebras has a state of maximum entropy. Indeed we have seen that, at least compared to other black hole solutions, empty dS space is the solution with maximum entropy, see section \ref{sec:ds_entropy}. All other black hole solutions will have less entropy than the empty dS one. This maximum entropy state will have a density matrix $\rho$ that can be normalised so that $\rho = \mathds{1}$. Then, all entanglement eigenvalues are equal, so it is said to have a ``flat entanglement spectrum'' or to be a ``maximally mixed'' state. It can also be thought as a thermal state at infinite temperature, $\beta \to 0$. These properties hold perturbatively in $G_N$.

A pressing question is how to incorporate these abstract ideas into a more standard holographic framework. It turns out that the answers might depend on the spacetime dimension under consideration. In the following, we will consider different approaches to static patch holography, starting with some ideas for higher dimensional gravity and then, discussing some particular proposals in $d=3,2$.

\subsubsection{$d\geq4$: conformal boundary conditions}

An initial strategy to do holography inside the static patch could be to put an \textit{artificial} boundary à la York and describe the system in terms of the now fixed boundary data. As mentioned, inside the static patch this problem was studied in \cite{Hayward:1990zm, Wang:2001gt}. Recently, it has been thoroughly extended (including SdS black holes) in \cite{Draper:2022ofa, Banihashemi:2022jys}.

Quite remarkably, in parallel to these developments, there have been recent advances in the mathematical relativity community involving the Initial Boundary Value Problem (IBVP) in General Relativity \cite{Witten:2018lgb, An:2021fcq}. Consider a 
$d$-dimensional manifold of the form $\mathcal{M} = I \times S$, with $I$ being a finite interval and $S$, a spatial $(d-1)$-manifold with non-empty boundary $\partial S  = \Sigma$. The boundary of $\mathcal{M}$ is denoted $\mathcal{C} = I \times \Sigma$. Now the IBVP consists on finding metrics on $\mathcal{M}$ that are solutions to the vacuum Einstein equations,
\begin{equation}
    R_{\mu\nu} - \frac{1}{2} R \, g_{\mu\nu} + \Lambda \, g_{\mu\nu} = 0 \,,
\end{equation}
together with boundary conditions along $\mathcal{C}$ and initial conditions along a Cauchy surface, say at some initial slice. See \cite{Sarbach:2012pr} for a review on the subject. Here, it is important to have dynamical gravity, so $d\geq 4$. The inclusion (or not) of a cosmological constant does not change the results we will be reviewing below.

The statement in \cite{An:2021fcq} is that with respect to either Dirichlet (fixing the induced metric on $\mathcal{C}$) or Neumann (fixing the second fundamental form of $\mathcal{C}$ in $\mathcal{M}$) the IBVP for General Relativity is \textbf{not} well-posed. 

This statement comes in the form of two theorems. First, it is proven that given a generic set of boundary data, the constraint of Einstein equations at the boundary will not be satisfied (this is a gauge-independent proposition). Second, given a consistent set of boundary data and a solution which solves the IBVP, then there exist other infinitely many physically distinguishable solutions on $\mathcal{M}$ (this is proven in the harmonic, or de Donder gauge). The Euclidean analogue of this problem is reviewed in \cite{Witten:2018lgb}, with similar conclusions.

Note this is unlike other IBVPs that are known to be well-posed. Examples include scalar, Maxwell or Yang-Mills theory. It is also contrary to the Initial Value (or Cauchy) Problem in General Relativity that is known to be well-posed since the  60's \cite{1952AcMa...88..141F, Choquet-Bruhat:1969ywq}. 

One should find this problematic, specially if one wants to give a thermodynamic interpretation to the (Euclidean) gravitational path integral. Fortunately, there is another set of geometric data that is conjectured (not proven) to be well-posed. These are the so-called \textbf{conformal boundary conditions}, that are mixed conditions that involve fixing
\begin{equation}
    \text{Conformal boundary conditions:} \quad \{ \left[ h\right], K \} \,,
\end{equation}
where $\left[ h\right]$ is the conformal class of boundary metrics and $K$ is the trace of the extrinsic curvature at the boundary. There is a proof of the IBVP with conformal boundary conditions being well-posed in Euclidean signature \cite{Witten:2018lgb} and evidence for it in Lorentzian signature \cite{An:2021fcq, Anninos:2022ujl}.

It seems essential to understand the gravitational path integral in terms of conformal boundary data to make sense not only of static patch holography but also, of some other finite versions of holography, such as finite cutoff AdS \cite{Taylor:2018xcy, Hartman:2018tkw}. Finally, it is interesting to note that these conformal boundary conditions have been anticipated in \cite{Bredberg:2011xw, Anninos:2011zn}, where obstructions found with standard Dirichlet boundary conditions were bypassed by imposing conformal boundary conditions instead. 

\subsubsection{$d=3$: $T\bar{T} + \Lambda_2$ deformations}

The theorems discussed in the previous section require dynamical gravity in the bulk. In lower dimensions, there have also been interesting recent developments regarding timelike boundaries in dS. In particular, in $d=3$, there have been recent constructions involving certain solvable irrelevant deformations of CFT$_2$ that allow us to count the dS entropy, including its leading logarithmic correction, from certain AdS black hole microstates \cite{Shyam:2021ciy, Coleman:2021nor}.

The idea is to start from some seed CFT and deform it by two types of irrelevant deformations. At any step in the process, the Euclidean theory is defined through a differential equation,

\begin{equation}
\frac{\partial}{\partial \lambda} \log Z(\lambda, g) = -2\pi \int d^2 x \sqrt{g} \langle T \bar{T} \rangle + \frac{1-\eta}{2\pi \lambda^2} \int d^2 x \sqrt{g} \,.
\end{equation}

Here $Z(\lambda, g)$ is the partition function of the theory and depends on the background metric $g$ and the deformation parameter $\lambda$. $T\bar{T}$ is a quadratic combination of the stress tensor defined by $T\bar{T} \equiv \frac{1}{8} \left(T_{ab}T^{ab} - (T_a^a)^2 \right)$ \cite{Smirnov:2016lqw, Cavaglia:2016oda}. The last term corresponds to changing the cosmological constant in two dimensions and is usually called a $\Lambda_2$ deformation.

The idea is to start the process with some (holographic) CFT, so the appropriate boundary condition will be $Z_{\lambda=0, \eta=1} = Z_{CFT}$. We focus on the energy levels around $\Delta \approx c/6$, with $c \gg 1$ being the central charge of the CFT. Now the process is taken in steps. First, we deform the CFT with a pure $T\bar{T}$ deformation ($\eta = 1$). At some finite value of $\lambda = \lambda_0$, we turn on the $\Lambda_2$ deformation by setting $\eta = -1$. This deformation is solvable and it is possible to compute the energy spectrum at each step of the process.

\begin{figure}[h!]
        \centering
        \includegraphics[height=4cm]{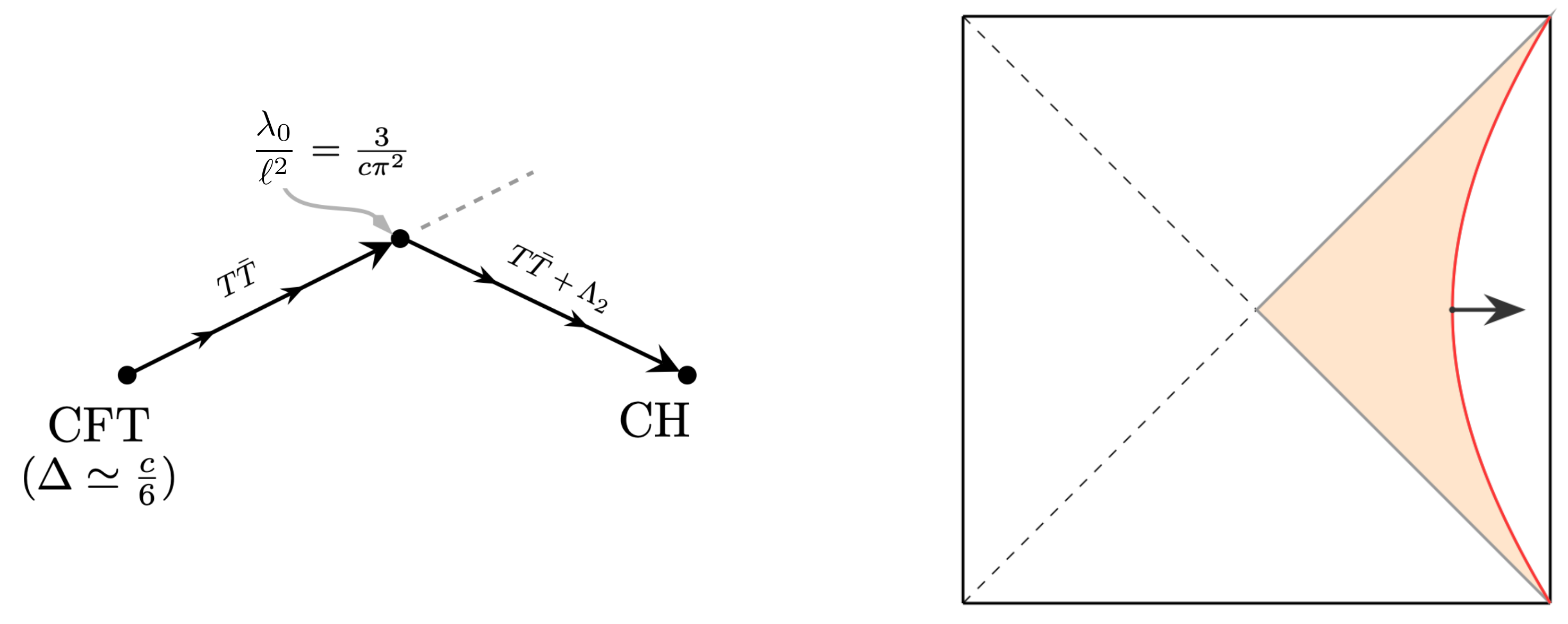}
\caption{Sketch of the $T\bar{T} + \Lambda_2$ deformation on the field theory (left) and on the Lorentzian gravity (right) side. Figure adapted from \cite{Coleman:2021nor}.}\label{fig:ttbar}
\end{figure}

On the gravity side, the dual of this process is as follows. We start with a BTZ black hole. The $T\bar{T}$ deformation brings the AdS boundary very close to the black hole horizon. At this point, the black hole horizon becomes indistinguishable from a cosmological horizon with the same radius. This is when the second $T\bar{T} + \Lambda_2$ deformation is turned on, building the static patch geometry in dS, from the horizon, up to a timelike boundary. This is called the cosmic horizon (CH) patch. See figure \ref{fig:ttbar}.

Through this process, the dressed black hole states can be shown to contribute to the dS entropy, obtaining that
\begin{equation}
S_{T\bar{T}+\Lambda_2} = S_{GH} - 3 \log (S_{GH}) + \cdots \,,
\end{equation}
which, remarkably, reproduces not only the leading Gibbons-Hawking entropy but also the logarithmic correction found in \cite{Anninos:2020hfj}.

Initially, this construction was proposed in the context of the dS/dS correspondence \cite{Alishahiha:2004md, Gorbenko:2018oov}, where the dS patch was reconstructed. See section \ref{others}. It is also possible to start with states around $\Delta = 0$ and in that case, the construction does not reproduce the Gibbons-Hawking entropy. Instead, in the gravity picture, it describes a region with no horizon, that goes from the timelike boundary towards the observer's worldline. Maybe a combination of both approaches might provide a full global dS picture \cite{Coleman:2021nor}. Moreover, recently the possibility of embedding this formalism in string theory has been explored \cite{Silverstein:2022dfj}. The $T\bar{T}$ deformation has also been used with potential applications to the dS/CFT correspondence \cite{Araujo-Regado:2022gvw}. Finally, it would be interesting to include matter in these deformed theories and try to probe some dynamical features of the cosmological horizon, such as the Gao-Wald effect, as a non-trivial check of this proposal.

\subsubsection{$d=2$: open quantum mechanical duals}
Two dimensions is yet another avenue to explore holographic features of dS, as they offer some extra advantages compared to higher dimensions. First, following the standard holographic dictionary, the dual theory should be a (maybe disordered) quantum mechanical theory, which is easier to deal with than the usual holographic quantum field theories. Moreover, from the gravity perspective, there are dilaton-gravity theories that admit solutions where the cosmological horizon is in causal contact with an AdS boundary, where the observer is under control. In higher dimensions, this would be forbidden by the null energy condition \cite{Freivogel:2005qh}. 

The action for a generic dilaton-gravity (Euclidean) theory in two dimensions is given by
\begin{equation}
I_E = - \frac{1}{16\pi G_N} \int_{\mathcal{M}} d^2x \sqrt{g}\left( \Phi R + U(\Phi)  \right) - \frac{1}{8\pi G_N} \int_{\partial\mathcal{M}} du \sqrt{h} \Phi_b K~,
\end{equation}
where $\Phi$ is the dilaton field, and $\Phi_b$ is the value of $\Phi$ at $\partial\mathcal{M}$. $U(\Phi)$ is usually called the dilaton potential. We have chosen a frame in which the action contains no $\Phi$ derivatives.\footnote{Other choices containing terms with derivatives of the dilaton that also admit solutions with dS interiors can be found in, for instance, \cite{Grumiller:2021cwg, Ecker:2022vkr}.} It is easy to see that solutions to this action obey $R = -U'(\Phi)$, so the usual JT gravity is the particular case where $U(\Phi) = 2\Phi$, and dS can be obtained whenever $U(\Phi) = -2\Phi$.  Quasi-local thermodynamics in dS JT gravity have been recently explored in \cite{Svesko:2022txo}. 

Geometries that interpolate between a dS horizon and an AdS boundary can be obtained as solutions to smooth potentials that asymptotically behave as $U(\Phi) = 2|\Phi|$. These are usually called centaur geometries \cite{Anninos:2017hhn, Anninos:2018svg}, and allow us to probe the cosmological horizon using the standard tools of AdS holography. More generic forms of $U(\Phi)$ have been extensively studied in, for instance, \cite{Cavaglia:1998xj, Grumiller:2007ju,Witten:2020ert}.

It is clear, then, that the problem of quantum dS in two dimensions becomes the problem of how to microscopically modify the dilaton potential $U(\Phi)$ away from AdS. One possibility is to use the equivalence of JT gravity with a matrix model \cite{Saad:2019lba}. The dilaton potential can be deformed away from linear with the inclusion of conical defects \cite{Maxfield:2020ale,Witten:2020wvy, Turiaci:2020fjj, Eberhardt:2023rzz}. It seems plausible that this construction might be able to accommodate at least a finite piece of dS spacetime \cite{Anninos:2022hqo}.

Another model that has resemblance to JT gravity at low energies is the Sachdev-Ye-Kitaev (SYK) model \cite{Sarosi:2017ykf, Trunin:2020vwy}. Following the idea of holographic renormalisation group flow \cite{deBoer:1999tgo}, deformations away from AdS$_2$ in the gravity picture would correspond to relevant deformations in the quantum side. Interestingly, the SYK model has a large set of tractable relevant deformations whose Hamiltonian takes the form \cite{Garcia-Garcia:2017bkg, Jiang:2019pam, Lunkin:2020tbq, Anninos:2020cwo, Nandy:2022hcm, Anninos:2022qgy, Khveshchenko:2023upm},
\begin{equation}
H = H_{\text{SYK}}^q + \sum_i \lambda_i \, H_{\text{SYK}}^{\tilde{q}_i} \,.
\end{equation}
Here the full Hamiltonian is the sum of multiple SYK Hamiltonians with a different number $q, \tilde{q}_i$ of fermions in the all-to-all interactions. If $\tilde{q}_i<q$, then the second term acts as a relevant deformation that is controlled by the dimensionless couplings $\lambda_i$. Even though unitarity requires $\lambda_i \in \mathbb{R}$, there is evidence that allowing for complex couplings might give rise to macroscopic features consistent with dS space \cite{Anninos:2020cwo, Anninos:2022qgy}. If this is correct, it would imply that the microscopic theory for the static patch might be non-unitary and should be treated as a holographic open quantum system. We have seen during these lecture notes some behaviour compatible with this. See, for instance, the shockwave solution in chapter \ref{chap:shocks}. Recent progress in understanding non-unitary generalisations of the SYK model can be found in \cite{Liu:2020fbd, Garcia-Garcia:2021elz, Zhang:2021klq,Garcia-Garcia:2021rle, Sa:2021tdr, Kulkarni:2021gtt, Kawabata:2022osw, Garcia-Garcia:2022rtg, Kawabata:2022cpr}. As conformal symmetry in the SYK model is only emergent in the large $N$ limit, it is tempting to think that dS unitarity might also be an emergent property from the microscopic theory.

\vspace{0.3cm}

\begin{center}  \rule{0.5 \textwidth}{0.5pt} \end{center}

\vspace{0.3cm}

Given all the recent developments and new ideas discussed in this Chapter, we would like to finish these notes by updating exercise $4$ in \cite{Spradlin:2001pw} into a form that is still quite challenging but hopefully more achievable these days:
\begin{empheq}[box={\mymath[drop lifted shadow]}]{gather*}
\notag\text{Independently of the number of dimensions, matter content, or type of holography,}
\\ \notag\text{find a microscopic quantum theory dual to, at least, a patch of de Sitter space.}
\end{empheq}

\hspace{1cm}

\acknowledgments
\noindent I would like to thank all the organisers and participants of the XVIII Modave Summer School in Mathematical Physics for a very stimulating environment. I would also like to thank D. Anninos, S. Chapman, E. Harris, C. Maneerat, E. \'O Colg\'ain, D. Pardo Santos, B. Pethybridge, A. Rios Fukelman, E. Shaghoulian, and S. Sheorey for interesting discussions and feedback on the manuscript. Special thanks to C. Maneerat for creating most of the plots presented in these notes.  I acknowledge valuable discussions with participants from the workshop "Quantum de Sitter Universe" held at DAMTP, Cambridge and funded by the Gravity Theory Trust and the Centre for Theoretical Cosmology. My work is funded by UKRI Stephen Hawking Research Fellowship ``Quantum Emergence of an Expanding Universe".

\appendix

\section{Black hole entropy}
\label{sec:bh_entropy}
In this appendix, we provide more details on the computation of the black hole entropy from the Euclidean path integral, see box \ref{prf:bhent}. The reason to do so is twofold. On the one hand, it is to contrast the difference in the calculation of the entropy of a black hole and a cosmological horizon. In the latter, there is no need to include either boundary terms or regularisation to compute the path integral at the dS saddle point. On the other, we will show the use of the York timelike boundary in the computation as a way of properly defining the thermodynamical ensemble. Similar ideas might play a role in dS holography as described in section \ref{sec:holo_ds}.

But let's go back to the calculation. We are in Euclidean signature in four dimensions and we want to evaluate the path integral,
\begin{equation}
Z = \int Dg \exp (-I_{E}[g]) \quad , \quad I_E = - \frac{1}{16\pi G_N}\int d^4x \sqrt{g} \, R - \frac{1}{8\pi G_N} \int_{r=r_0} d^3x \sqrt{h} \, K \,,
\end{equation}
in the saddle-point approximation as $G_N \to 0$. We already included a boundary term at a fixed $r=r_0$, from which we will specify the boundary data. One saddle point is the Euclidean black hole,
\begin{equation}
    ds^2 = \left(1- \frac{2M}{r}\right) dt_E^2 + \frac{dr^2}{\left(1- \frac{2M}{r}\right)} + r^2 d\Omega_2^2 \,, \label{sch_bh_app}
\end{equation}
that, as discussed, has $t_E \sim t_E + 8\pi M$. As in \cite{York:1986it}, we will not directly identify $8\pi M$ with the inverse temperature. Instead we will specify our thermodynamical variables at the boundary. In this case, the boundary at $r=r_0$ is $S^1 \times S^2$, so we will use the proper length of the $S^1$ and the area of the $S^2$ as our independent variables,
\begin{equation}
     \left\{\begin{array}{l}
        \beta (r_0) = \int_0^{8\pi M} d\tau_E \left(1-\frac{2M}{r_0}\right)^{1/2} = 8\pi M \left(1-\frac{2M}{r_0}\right)^{1/2}, \label{74} \\
       A(r_0) = \int d\Omega_2 r_0^2 = 4\pi r_0^2 \,.
        \end{array}\right.
\end{equation}
Now $\beta(r_0)$ is the local temperature that an observer feels at $r_0$. Note that given $\beta(r_0)$, there might be more than one positive $M$ that satisfies (\ref{74}), or even no positive solution. As $r_0 \to \infty$, we recover the usual statement that $\beta_\infty = 8\pi M$.

The bulk action is zero but the boundary term is non-vanishing. Evaluating the action on the solution (\ref{sch_bh_app}) gives the on-shell action,
\begin{equation}
    I_E^{\text{on-shell}} = \frac{12 \pi}{G_N} M^2 - \frac{8\pi}{G_N} M \, r_0 \,, 
\end{equation}
that diverges as $r_0 \to \infty$. The way of regulating the computation is by subtracting a boundary counterterm in the action. What York proposes \cite{York:1986it} is to subtract the same action $I_E$, but now evaluated on a \textit{flat} metric with the same $S^1 \times S^2$ boundary as in the black hole metric,
\begin{equation}
    ds^2_{\text{subtract}} = d\tau^2 + dr^2 + r^2 d\Omega_2^2 \,,
\end{equation}
with $\tau \sim \tau + \beta$. Again the bulk action vanishes and using that $K = -2/r_0$ at the boundary, we obtain
\begin{equation}
    I_E^{\text{subtract}}= - \frac{\beta (r_0) r_0}{G_N} = - \frac{8\pi}{G_N} M r_0 \left(1-\frac{2M}{r_0}\right)^{1/2}\,.
\end{equation}
Then,
\begin{equation}
    I_E^{\text{finite}} = I_E^{\text{on-shell}} -  I_E^{\text{subtract}} = \frac{12 \pi}{G_N} M^2 - \frac{8\pi}{G_N} M r_0 \left( 1- \left(1-\frac{2M}{r_0}\right)^{1/2} \right) \,,
\end{equation}
which is now finite as $r_0\to \infty$. For general metrics, this regularisation procedure is not completely understood, except in the context of asymptotically AdS spacetimes, where a more sophisticated version of counterterm subtraction goes under the name of holographic renormalisation \cite{Skenderis:2002wp}.

In order to compute thermodynamic quantities, we need to write $I_E^{\text{finite}}$ not in terms of $M$, but in terms of $\beta$, while keeping the area $A$ fixed. We identify $I_E^{\text{finite}}$ as $\beta F$, the free energy of the system. Then, in order to compute the entropy one needs to do,
\begin{equation}
    S = \left. \left(1- \beta \partial_\beta \right)\right|_A (-I_E^{\text{finite}}) = \frac{4\pi M^2}{G_N} = \frac{A_H}{4G_N} \,,
\end{equation}
which is exactly the Bekenstein-Hawking entropy for the black hole. Note, first, that as opposed to the Gibbons-Hawking calculation of the dS entropy, here the area term comes from a boundary contribution to the action. Second, even though we needed regularisation of the free energy, given that the counterterm was proportional to $\beta$, it does not contribute to the entropy.

Finally, we can also compute the specific heat,
\begin{equation}
    C = \left. \left( \beta^2 \partial_\beta^2 \right) \right|_A (-I_E^{\text{finite}}) = -\frac{8\pi M^2}{G_N} \frac{1-\frac{2M}{r_0}}{1-\frac{3M}{r_0}} \,,
\end{equation}
which, interestingly, can be positive for $2M < r_0 < 3M$, in contrast to the negative specific heat in the standard calculation when $r_0 \to \infty$. This is an interesting effect that appears in the presence of finite timelike boundaries. As discussed in section \ref{sec:holo_ds}, it has been recently proven that fixing the induced metric on the boundary does not lead to a well-posed problem \cite{An:2021fcq}, so it would be desirable to revisit this calculation in the light of the new conjectured boundary conditions that do lead to well-posedness, \ie fixing the conformal class of the metric at the boundary and the trace of the extrinsic curvature.

\bibliographystyle{JHEP}

\bibliography{bibliography}

\end{document}